\documentclass[%
reprint,
amsmath,amssymb,
aps,
 prl,
floatfix,
twocolumn]{revtex4-2}

\usepackage{graphicx}
\usepackage{dcolumn}
\usepackage{bm}
\usepackage{amsmath} 
\usepackage{amssymb}
\usepackage{mathrsfs}
\usepackage{siunitx}
\usepackage{xcolor}
\usepackage{tabularx}
\usepackage{microtype}
\usepackage{graphicx,wrapfig,lipsum}

\newcommand{\fs}[1]{{#1}} 

\begin{document}

\title{Available observation time regulates optimal balance between sensitivity and confidence}

\author{Sahel Azizpour$^{1,2}$}
\author{Viola Priesemann$^{3,4}$}
\author{Johannes Zierenberg$^{3,*}$}
\author{Anna Levina$^{1,2,*}$}

\affiliation{
  $^1$\mbox{Eberhard Karls University of Tübingen, T{\"u}bingen, Germany}
  $^2$\mbox{Max Planck Institute for Biological Cybernetics, T{\"u}bingen, Germany}
  $^3$\mbox{Max Planck Institute for Dynamics and Self-Organization, Am Fa\ss berg 17, 37077 G{\"o}ttingen, Germany},\\
  $^4$\mbox{Institute for the Dynamics of Complex Systems, University of G\"ottingen, Friedrich-Hund-Platz 1, 37077 G\"ottingen, Germany},\\
  $^*${These authors contributed equally}
}

\date{\today}

\begin{abstract}
Tasks that require information about the world imply a trade-off between the time spent on observation and the variance of the response.
In particular, fast decisions need to rely on uncertain information.
However, standard estimates of information processing capabilities, such as the dynamic range, are defined based on mean values that assume infinite observation times. 
Here, we show that limiting the observation time results in distributions of responses whose variance increases with the temporal correlations in a system and, importantly, affects a system's confidence in distinguishing inputs and thereby making decisions.
To quantify the ability to distinguish features of an input, we propose several measures and demonstrate them on the prime example of a recurrent neural network that represents an input rate by a response firing averaged over a finite observation time.
We show analytically and in simulations that the optimal tuning of the network depends on the available observation time, implying that tasks require a ``useful'' rather than maximal sensitivity.
Interestingly, this shifts the optimal dynamic regime from critical to subcritical for finite observation times and highlights the importance of incorporating the finite observation times concept in future studies of information processing capabilities in a principled manner.
\end{abstract}

\maketitle

\section{Introduction}
\fs{Interacting with an environment requires agents to process information in a finite amount of time.} 
In general, the available amount of time will determine the certainty of the processed information underlying a decision.
While short times may suffice to interpret simple information, longer times are required to capture more complex information~\cite{hogarth_decision_1975}. 
As such, the available observation time determines the complexity of achievable tasks, and, in reverse, information processing optimization should be tailored to the available observation time.

\fs{One particularly fascinating system with enormous information processing capability is the brain.}
While the complex interplay of different brain areas is crucial for behavioral responses, local neuronal circuits were shown to encode specific information about the input. 
For example, the average neural firing rate recorded from neurons in monkey visual cortex (MT) was shown to encode the correlation of moving dots in the visual field, largely consistent with the behavioral decision~\cite{britten_analysis_1992}.
Similarly, single-neuron firing responses were shown to encode odor concentration~\cite{wachowiak_representation_2001} and sound level~\cite{evans_dynamic_1981,dean_neural_2005}.
These examples demonstrate a common leading-order approach to quantify a system's ability to encode inputs with stationary responses in a so-called neural response curve that relates some input feature (e.g., the stimulus intensity) to some output feature (e.g., the mean firing rate).
The encoding capacity can then be quantified by the dynamic range, i.e., the range of inputs that is mapped to a reasonable response range~\cite{kinouchi_optimal_2006, gautam2015maximizing, zierenberg_tailored_2020}.

If, however, one only has a finite time to integrate over responses, then the ability to distinguish features of the input will be affected by the intrinsic noise of the system.
Proposed concepts to encode input in the presence of noise and variability range from attractor dynamics~\cite{hopfield_neural_1982,battista_capacity-resolution_2020}, to stable transient dynamics with metastable states~\cite{rabinovich_transient_2008}, to low-dimensional manifolds in the full population response~\cite{gallego_neural_2017, vyas_computation_2020}.
All these concepts have in common that --- for finite observation times --- responses will vary from observation to observation, raising the question how to quantify the ability to distinguish inputs in a principled manner.

{In this work, we study how a finite observation time affects the ability to discriminate inputs from the response of a stochastic system.}
To quantify the ability to distinguish inputs, we develop suitable measures based on the distribution of responses instead of their mean.
We find that -- depending on the available observation time -- different dynamical regimes become more suitable for discrimination. 
In particular, for an infinite observation time the optimal regime is at a critical point, while for finite observation times the optimal regime shifts to subcritical states.
While this finding can be intuitively explained by the increasing sensitivity and variability (and hence uncertainty) when approaching a critical point, it adds new arguments to the ongoing debate about optimal information processing capacities of (near)critical systems~\cite{beggs_criticality_2008, chialvo_emergent_2010, munoz_colloquium_2018, wilting_operating_2018, wilting_25_2019}.

\begin{figure*}[t]
    \centering
    \includegraphics[width = 1\textwidth]{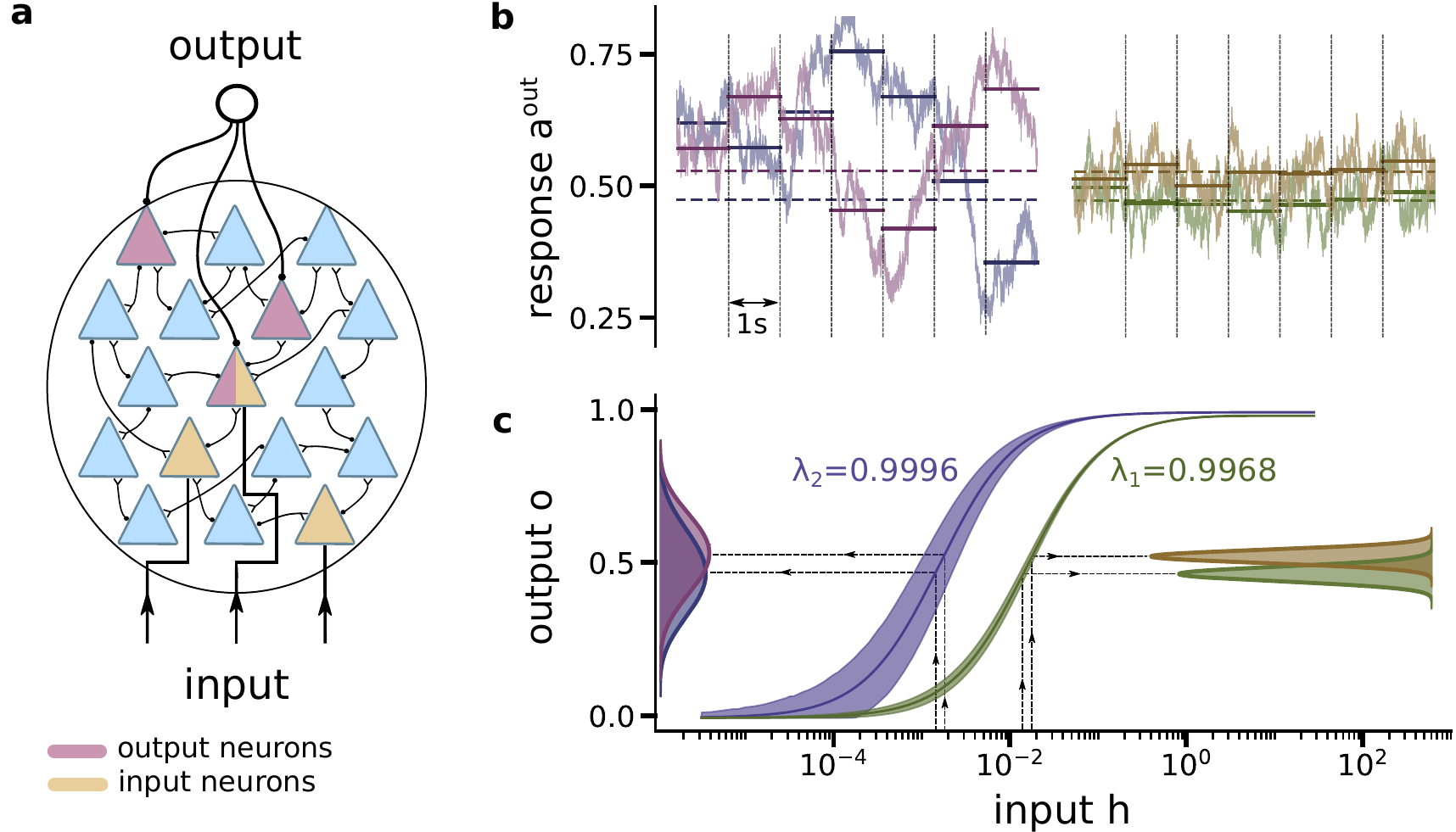}
    \caption{
    \textbf{Fluctuations in network activity on timescales larger than the observation window lead to the unreliability of input reconstruction.}
    \textbf{a)} Illustration of the recurrent neural network.
    \textbf{b)} Temporal evolution of outputs for two networks with different dynamical states $\lambda_1 = 0.9968$ (left, purple) $<$ $\lambda_2 = 0.9996$ (right, yellow) presented with two different inputs.
    The inputs are chosen such that the infinite time mean output rates are equal in both networks (dashed lines).
\textbf{c)} The mean output rate over a finite observation time as a function of the input rate for two networks with different dynamical states as in (b). The solid lines shows the asymptotic mean for infinite observation time, shading indicate variability of inferred mean over many observation windows of length $T=1s$.The projected probability distributions of outputs for two neighboring input values indicate that the distributions have larger overlap when the network is closer to criticality (overlap area $0.82$ vs. $0.21$).
    }
    \label{fig1}
\end{figure*}

\fs{While our considerations apply to general systems that process information about an input via stochastic output responses, we here focus on a particular example of a network of stochastic neurons that can be treated both numerically and analytically.}
Specifically, we consider a random network of probabilistic integrate-and-fire neurons:
In each time step, a neuron can be activated ($s^i_t=1$) recurrently or externally (see Methods for details). 
In brief, recurrent activation occurs with probability proportional to the weighted sum of active pre-synaptic neurons. 
Each neuron is connected to $K$ randomly selected neurons with identical weights $\lambda/K$ for mathematical tractability (see Supplemental Material for controls using Erdős–Rényi networks). 
External activation is modelled as direct activation from a binary input signal, which in our examples are independent Poisson processes with rate $h$.
To mimic information processing and transmission through the cortical hierarchy, only a subset of $N^\mathrm{in}$ input neurons receives input and we will ask how well this input is encoded by the response $a^\mathrm{out}_t=\sum_{i=1}^{N^\mathrm{out}} s^{\mathrm{out},i}_t/N^\mathrm{out}$ of another subset of $N^\mathrm{out}$ output neurons (see Fig.~\ref{fig1}a for an illustration, details of the model in Methods). 
As in biological systems, the network has a finite number of time steps $T$ to generate an output $o_t^T=\frac{1}{T}\sum_{i=0}^{T} a^\mathrm{out}_{t-i}$.
The task of the network is thus to reliably transmit the incoming signal $h$ from the input population to produce an informative output  within a time interval $T$ for further processing.

\fs{In this model, the network dynamics are controlled by the largest eigenvalue $\lambda$ of the connectivity matrix~\cite{larremore2011predicting}}
Without external input, the network has a critical point at $\lambda_c=1$, which is argued to optimize multiple information processing properties~\cite{munoz_colloquium_2018}.
At the same time, however, $\lambda$ controls the magnitude of temporal fluctuations in activity: if $\lambda$ approaches the critical point  ($\lambda\to 1$ for our model), network fluctuations increase; if $\lambda$ decreases to zero, neurons become isolated, and network fluctuations vanish (Fig.~\ref{fig1}b). 
This generates a trade-off: close to criticality, one expects optimized information processing properties for infinite observation time but simultaneously increased fluctuations in the finite-time output $o_t^T$. 
In the following, we explore how to solve this trade-off depending on the available observation time. 

\fs{As a first step, we illustrate how the networks that differ only a little in connectivity parameters can have very different encoding stability. }
We consider two networks with $\lambda_1=0.9968 < \lambda_2 = 0.9996$ and, for a fair comparison, chose input pairs that produce matching mean output signals (Fig.~\ref{fig1}c)
Moderate fluctuations in the first network allow visually distinguishing the two inputs using the output signal from finite observation times (Fig.~\ref{fig1}b, right: yellow bars reliably higher than green bars).
However, due to larger fluctuations in the second network, individual estimates from finite observation times are no longer ordered, and the two inputs can no longer be discriminated (Fig.~\ref{fig1}b, left: magenta and purple bars change order).

\fs{To formalize the intuition gained from the example, we consider the input-output distribution $P(\hat{o}|h)$ to observe the output $\hat{o}$ in response to a specific input $h$.}
In our previous case, $\hat{o} = o_t^T$, where variability comes from observing stochastic, correlated dynamics for a finite amount of time, but the logic remains the same for other causes of variability. 
Now, if two inputs $h_1$ and $h_2$ were equally likely to be presented, then the overlap between $P(\hat{o}|h_1)$ and $P(\hat{o}|h_2)$ quantifies the minimal discrimination error~\cite{berens_neurometric_2009} of an ideal observer:
\begin{equation}
\label{Eq:MDE}
    \mathcal{E}(h_1,h_2)=\frac{1}{2}\int \min\left\{P(\hat{o}|h_1),P(\hat{o}|h_2)\right\}d\hat{o}.
\end{equation}
Computing this error for the stimuli in our example (Fig.~\ref{fig1}c), we obtain for $\lambda_1$ an error $\mathcal{E}_1 = 0.10$, and for $\lambda_2$ a much larger error $\mathcal{E}_2 = 0.41 > \mathcal{E}_1$.

\begin{figure}
    \includegraphics[width=0.48\textwidth]{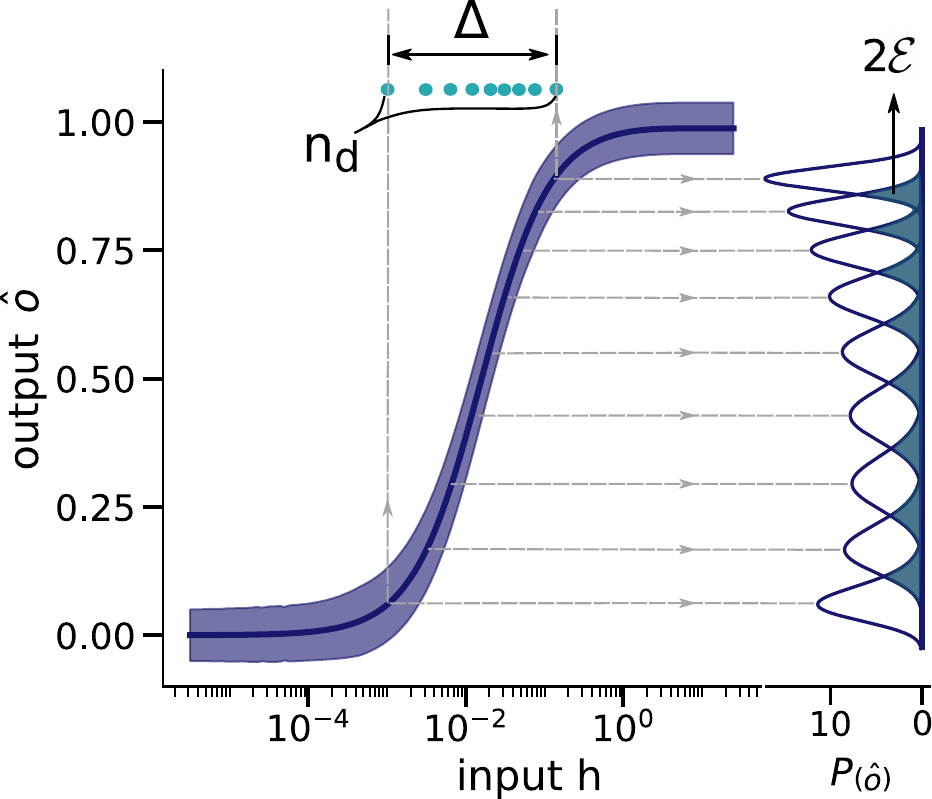}
     \caption{\fs{Definition of the variability-aware dynamic range.} 
     Input intensities indicated on top in cyan can be distinguished with the minimal discrimination error $\mathcal{E} < \varepsilon$, as seen from the overlap if the noisy output distributions (indicated on the right). For each error-threshold $\varepsilon$ we can define the number $N_d$ and the range $\Delta$ of inputs that can be reliably discriminated. Here we used $\lambda=0.9968$, $\varepsilon=0.1$ and $\sigma_{noise}=0.02$ for descriptiveness of visualizations.}
    \label{fig2} 
\end{figure}

\begin{figure*}[t]
    \centering
    \includegraphics[width = 1\textwidth]{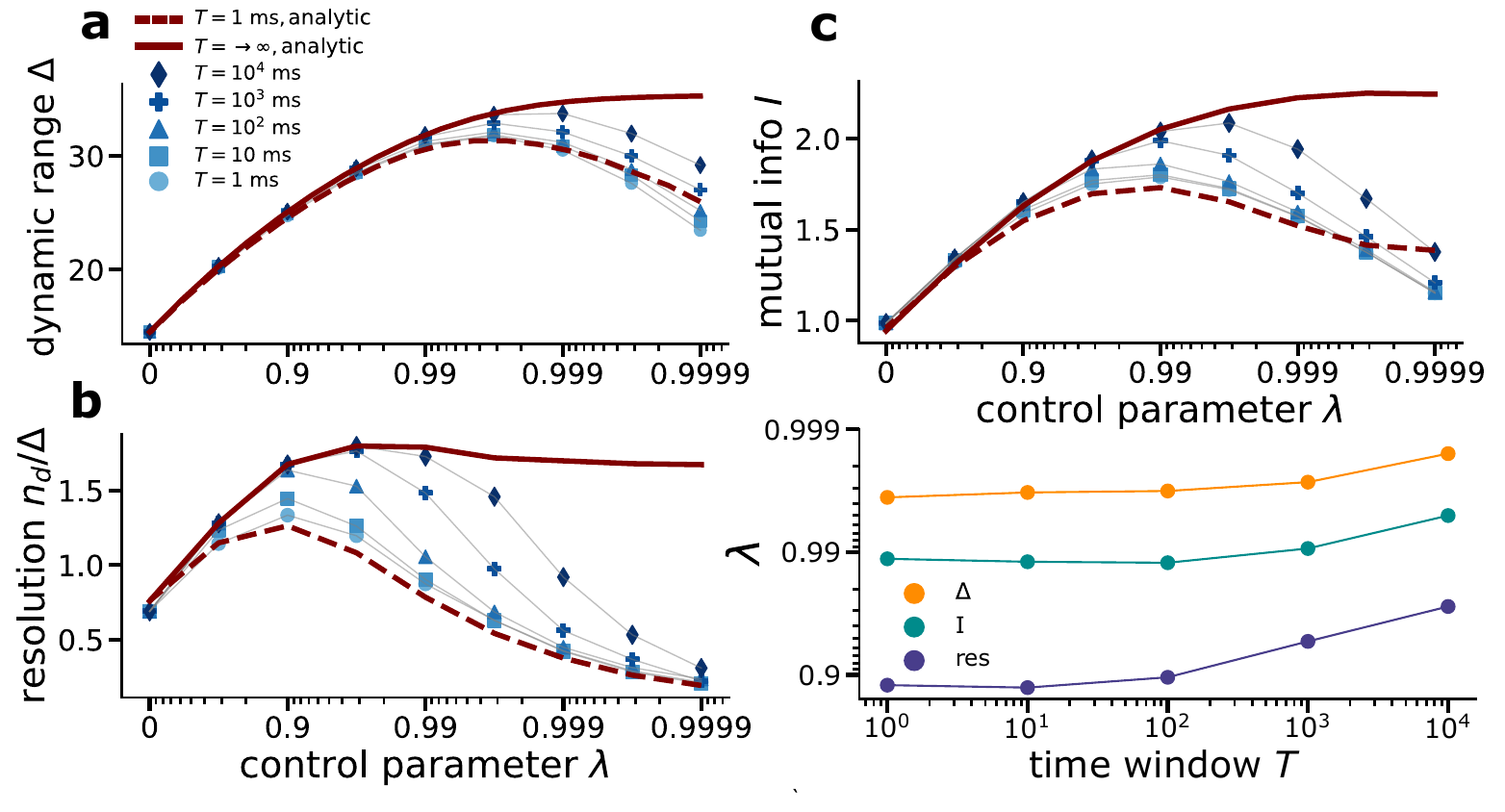}
    \caption{\fs{Dynamical regime for optimal information transmission depends on the observation time} \textbf{a)} Dynamic range, \textbf{b)} inference resolution and \textbf{c)} Mutual information between input and mean output activity given different observation times. Parameter values are $N^\mathrm{in}=2000$, $N^\mathrm{out}=10000$, $\varepsilon=0.2$ and $\sigma_{noise}=0.01$.
    \label{fig3}}
\end{figure*}

\fs{As a next step, we define a set of \emph{discriminable inputs} that can be sufficiently well distinguished from each other when observing only the output. }
We call two inputs $\varepsilon$-discriminable if the overlap of the response distributions generated by the inputs is smaller than an error-threshold $\varepsilon$.
Formally speaking, a set of $\varepsilon$-discriminable inputs $\mathcal{H}=\{h_1,h_2, \ldots, h_{n_d}\}$, with $h_0 = 0$ and $h_{n_d +1} = :h_\infty$, is a set for which $\mathcal{E}(h_i, h_j)\leq\varepsilon$ for all $i \neq j, \: i,j \in [0, n_d+1]$.
Finding the maximal (in the sense of cardinality) set of discriminable inputs is a close-packing problem without a unique solution. 
To circumvent this complication, we propose the following algorithm: start by finding $h_1^\mathrm{left} = \min\{h >h_0 = 0 : \mathcal{E}(h_0, h)\leq\varepsilon\}$, and then proceed by induction to $h^\mathrm{left}_{i+i} = \min \{h > h_i^\mathrm{left} : \mathcal{E}(h_i^\mathrm{left}, h)\leq\varepsilon \}$, (Fig.~\ref{fig2}, see Methods for more details). 
We stop at the first $i$ such that $\mathcal{E}(h_{i+1}^\mathrm{left},h_\infty) > \varepsilon$ and get this way $n_d^{\mathrm{left}} = i$.
We repeat the same procedure starting from the right with $h_1^\mathrm{right} = \max\{h < h_{\infty} : \mathcal{E}(h, h_\infty)\leq\varepsilon \}$ and iterate until $\mathcal{E}(h_{i+1}^\mathrm{right},0) > \varepsilon$ to find $n_d^\mathrm{right}$.
Our final estimate of discriminable inputs is then the average $n_d = 1/2 (n_d^{\mathrm{left}} + n_d^{\mathrm{right}})$.

While this algorithm is numerically straightforward, it comes with two technical challenges for our given example.
First, the output of our model is bounded by absorbing states such that in the limits $h\to 0$ and $h\to\infty$ the outputs become $\delta$-distributed, and overlap cannot be computed.
To circumvent this technicality, we further assume the output to be corrupted by Gaussian noise (e.g., due to readout noise) with zero mean and small variance $\sigma^2$ such that $\hat{o}^T_t = o_t^T + \sigma\eta$ with $\eta \sim \mathcal{N}(0,1)$.
The distribution of noisy outputs can then be computed as a convolution of the noise-free output distribution $P(o)$ and the Gaussian distribution $\mathcal{N}(0,\sigma^2)$, and for $\sigma \to 0$ we recover the noise-free version. 
Second, it takes a lot of compute time to obtain the exact values of $h^{\mathrm{left}/\mathrm{right}}_i$. 
We speed up the computations by introducing an analytical approximation, where all measured $P(o^T|h)$ are first fitted by a Beta distribution and parameters are interpolated for non-measured $h$ (see Methods and extended data figure Fig.~\ref{fig:extData_workflow}).

\fs{Using the set of $\varepsilon$-discriminable inputs we can now construct measures for information processing capabilities during finite observation intervals (Fig.~\ref{fig3})}. 
To compare to analytical calculations (see Methods) from mean-field responses for $T\to\infty$ (red solid lines) and $T\to 0$ (red dashed lines), we consider here the special case of $N^\mathrm{out}=N$, but our conclusions remain valid for $N^\mathrm{out}\ll N$ (see Extended Data Fig.~\ref{supp2}).

Let us start with the countable number of $\varepsilon$-discriminable points $n_d$ (Fig.~\ref{fig3}a).
One can see that $n_d$ first increases with $\lambda$ because the transmission task requires sufficient coupling between input and output population.
However, $n_d$ exhibits a maximum at a $T$-dependent subcritical $\lambda<1$, above which it decays presumably down to zero in the limit $\lambda\to 1$. 
Our numerical results interpolate between the analytical predictions for different observation times, indicating that every finite observation time will have an optimal $\lambda$, while for infinite observation time $n_d$ is bound by the Gaussian noise of the readout.

Let us now turn to the dynamic range, a measure that is commonly used in the field to asses coding capabilities~\cite{kinouchi2006optimal,munoz_colloquium_2018,shew_neuronal_2009}.
Recall that the dynamic range is typically defined as the logarithmic width of the input range for which the output $o^\infty(h)$ is between 10$^{\mathrm{th}}$ and 90$^{\mathrm{th}}$ percentile of all outputs~\cite{kinouchi_optimal_2006}, i.e., $\Delta=10\log_{10}(h_{0.9}/h_{0.1})$.
The arbitrary percentile thresholds are selected to cut off responses that would not be distinguishable from the noise floor at low activity and saturation regime at high activity. 
A natural generalization is thus to replace these thresholds with the first and last input that can be discriminated from the boundaries, defining the \textit{processing dynamic range}
\begin{equation}
    \Delta = 10\log_{10}(h_1^\mathrm{left}/h_1^\mathrm{right}),
\end{equation}
which depends on the specific choices of the discrimination error $\varepsilon$ and the Gaussian noise $\sigma$, which we tune in the following to recover the typical 10\%-90\% bounds of the established dynamic range~\cite{kinouchi_optimal_2006} for $T\to\infty$ (cf., Fig.~\ref{fig2}).
Thereby our processing dynamic range recovers the classical definition for $T\to\infty$ (Fig.~\ref{fig3}b).

For finite $T$, our numerical estimates interpolate well between our analytical bounds.
Importantly, a finite $T$ results in a strong reduction of the dynamic range in the vicinity of the critical point, i.e., for $\lambda\approx 1$, but only a small reduction at small $\lambda$. 
As a result, the finite-observation dynamic range develops a $T$-dependent maximum, which is, however, different from the maximum of $n_d$ (Fig.~\ref{fig3}d).

\fs{A complementary approach to quantify how fluctuations of the system affect its processing capacity is to use information theory.}
Specifically, we could ask ``what is the amount of information obtained about the input when observing the output?''.
This is quantified by the mutual information $I(\hat{o};h)$ between the estimated output variable $\hat{o}$ and the input variable $h$, a concept that has recently been applied successfully to study processing capacities of feed-forward deep neural networks~\cite{shwartz-ziv_opening_2017}.
The mutual information is calculated as $I(\hat{o};h) = H(\hat{o}) - H(\hat{o}|h)$, where $H(\hat{o})$ is the entropy of outputs and $H(\hat{o|}h)$ is a conditional entropy.
Formally, the (conditional) entropy for continuous variables is defined via integrals, however, since the activity in our model is discrete, we here proceed with $H(\hat{o})=-\sum_i P(\hat{o}_i)\log P(\hat{o}_i)$ as well as $H(\hat{o}|h)=-\sum_j P(h_j)\sum_i P(\hat{o}_i|h_j)\log P(\hat{o}_i|h_j)$.
The distribution of outputs can be estimated from all possible input-output distributions, $P(\hat{o})=\sum_k P(h_k)P(\hat{o}|h_k)$.
We now consider the most general case of a uniform distribution of inputs $P(h)=\frac{1}{n}$ to express the mutual information through $n$ samples of the output-response distribution.

\begin{align}
    I(\hat{o};h) = &\frac{1}{n}\sum_i\sum_j P(\hat{o}_i|h_j)\left[\log\ n\phantom{log\sum_k P(\hat{o}_i|h_k)}\right.\nonumber\\
    &\left.-\log\sum_k P(\hat{o}_i|h_k) + \log P(\hat{o}_i|h_j)\right].
\end{align}
Using the mutual information, we again find numerical estimates to interpolate between our analytical bounds (Fig.~\ref{fig3}c) with an optimal regime of $\lambda$ for information representation that depends on the available observation time (Fig.~\ref{fig3}d).

We thus find that, depending on which information processing quantity we aim to optimize, there are different optimal regimes (Fig.~\ref{fig3}d).
Moreover, even for biologically long observation times ($T>10^3\approx 1s$ for $\Delta t\approx 1ms$) all optima remain below $\lambda\approx0.999$.
It thus becomes a trade-off between different information processing capabilities, especially when noting that the number of $\varepsilon$-discriminable points practically vanish for $\lambda\to 1$ for all considered cases.

\section{Discussion}
To summarize, we find that, given a finite observation time, several measures of information processing capacities are maximal for subcritical dynamics ($\lambda<1$ in our model) and decrease when approaching the critical point.
The intuitive reason for this observation is that the emergent temporal fluctuations, which become stronger closer to a critical non-equilibrium phase transition, smear out the input signal and hinder discrimination.
This challenges the prevalent view in the field that systems have to be critical to optimize processing capacities~\cite{kinouchi_optimal_2006, beggs_criticality_2008, shew_neuronal_2009, shew_functional_2013, gautam_maximizing_2015}, but instead supports a more flexible view on how close-to-critical systems have freedom to optimize to the processing task~\cite{wilting_operating_2018, dahmen_strong_2022, zeraati2023intrinsic, khajehabdollahi2022critical}.

To illustrate how finite observation time affects information processing, we proposed several measures of input discriminability that can be generalized beyond neural networks.
These include the number of $\varepsilon$-discriminable points, an observation-time dynamic range, and mutual information.
All these measures are readily applicable to any scalar output that is informative about the input. 
However, we also showed that there are some systematic obstacles that need to be taken care of, in particular, related to absorbing phases, such that future steps include the derivation of more stringent measures. 
One promising way is to use information theory, as we have shown in the example of mutual information. 
While more abstract, it is also more fundamental, and one can make use of existing toolboxes to quantify information theoretic measures~\cite{lindner_trentool_2011, wollstadt_idtxl_2019}.

It should be noted that a single measure of processing ability, here to distinguish input, will likely not suffice to describe the performance of a system for a given task. 
For example, while we here observed for finite observation times a decreasing processing capacity as $\lambda\to 1$ for networks of probabilistic integrate-and-fire, the same networks actually shift their intervals of discriminable inputs with $\lambda$~\cite{zierenberg_tailored_2020}.
The increased fluctuations that are responsible for the decrease in dynamic range and resolution, thus enable to amplify otherwise undetectable input signals.
Such payoffs, e.g., an amplification that comes at the price of resolution, cannot be captured in scalar measures.
Moreover, it supports that a combination of networks with different dynamical states is a promising approach to efficient information processing~\cite{zierenberg_tailored_2020} in line with the hierarchical processing architecture of the primate cerebral cortex~\cite{felleman_distributed_1991}. 
Future developments of complex architectures that exploit this will require a better understanding of how network topology affects emergent collective dynamics, a problem that thrives for tools from physics.

To conclude, our results demonstrate that the optimal dynamical state of a system that processes information  will depend on the available observation time.
This is in line with recent results using neuromorphic chips for classification tasks, where more critical networks only outperform less critical networks for more complex tasks~\cite{cramer_control_2020}.
For the future, it will thus be necessary to specify the processing algorithm when making statements about which dynamical state is optimal.

\bibliography{bibliography}

\begin{thebibliography}{37}%
\makeatletter
\providecommand \@ifxundefined [1]{%
 \@ifx{#1\undefined}
}%
\providecommand \@ifnum [1]{%
 \ifnum #1\expandafter \@firstoftwo
 \else \expandafter \@secondoftwo
 \fi
}%
\providecommand \@ifx [1]{%
 \ifx #1\expandafter \@firstoftwo
 \else \expandafter \@secondoftwo
 \fi
}%
\providecommand \natexlab [1]{#1}%
\providecommand \enquote  [1]{``#1''}%
\providecommand \bibnamefont  [1]{#1}%
\providecommand \bibfnamefont [1]{#1}%
\providecommand \citenamefont [1]{#1}%
\providecommand \href@noop [0]{\@secondoftwo}%
\providecommand \href [0]{\begingroup \@sanitize@url \@href}%
\providecommand \@href[1]{\@@startlink{#1}\@@href}%
\providecommand \@@href[1]{\endgroup#1\@@endlink}%
\providecommand \@sanitize@url [0]{\catcode `\\12\catcode `\$12\catcode
  `\&12\catcode `\#12\catcode `\^12\catcode `\_12\catcode `\%12\relax}%
\providecommand \@@startlink[1]{}%
\providecommand \@@endlink[0]{}%
\providecommand \url  [0]{\begingroup\@sanitize@url \@url }%
\providecommand \@url [1]{\endgroup\@href {#1}{\urlprefix }}%
\providecommand \urlprefix  [0]{URL }%
\providecommand \Eprint [0]{\href }%
\providecommand \doibase [0]{https://doi.org/}%
\providecommand \selectlanguage [0]{\@gobble}%
\providecommand \bibinfo  [0]{\@secondoftwo}%
\providecommand \bibfield  [0]{\@secondoftwo}%
\providecommand \translation [1]{[#1]}%
\providecommand \BibitemOpen [0]{}%
\providecommand \bibitemStop [0]{}%
\providecommand \bibitemNoStop [0]{.\EOS\space}%
\providecommand \EOS [0]{\spacefactor3000\relax}%
\providecommand \BibitemShut  [1]{\csname bibitem#1\endcsname}%
\let\auto@bib@innerbib\@empty
\bibitem [{\citenamefont {Hogarth}(1975)}]{hogarth_decision_1975}%
  \BibitemOpen
  \bibfield  {author} {\bibinfo {author} {\bibfnamefont {R.~M.}\ \bibnamefont
  {Hogarth}},\ }\bibfield  {title} {\bibinfo {title} {Decision {{Time}} as a
  {{Function}} of {{Task Complexity}}},\ }in\ \href
  {https://doi.org/10.1007/978-94-010-1834-0_19} {\emph {\bibinfo {booktitle}
  {Utility, {{Probability}}, and {{Human Decision Making}}: {{Selected
  Proceedings}} of an {{Interdisciplinary Research Conference}}, {{Rome}},
  3\textendash 6 {{September}}, 1973}}},\ \bibinfo {series and number} {Theory
  and {{Decision Library}}},\ \bibinfo {editor} {edited by\ \bibinfo {editor}
  {\bibfnamefont {D.}~\bibnamefont {Wendt}}\ and\ \bibinfo {editor}
  {\bibfnamefont {C.}~\bibnamefont {Vlek}}}\ (\bibinfo  {publisher} {{Springer
  Netherlands}},\ \bibinfo {address} {{Dordrecht}},\ \bibinfo {year} {1975})\
  pp.\ \bibinfo {pages} {321--338}\BibitemShut {NoStop}%
\bibitem [{\citenamefont {Britten}\ \emph {et~al.}(1992)\citenamefont
  {Britten}, \citenamefont {Shadlen}, \citenamefont {Newsome},\ and\
  \citenamefont {Movshon}}]{britten_analysis_1992}%
  \BibitemOpen
  \bibfield  {author} {\bibinfo {author} {\bibfnamefont {K.}~\bibnamefont
  {Britten}}, \bibinfo {author} {\bibfnamefont {M.}~\bibnamefont {Shadlen}},
  \bibinfo {author} {\bibfnamefont {W.}~\bibnamefont {Newsome}},\ and\ \bibinfo
  {author} {\bibfnamefont {J.}~\bibnamefont {Movshon}},\ }\bibfield  {title}
  {\bibinfo {title} {The analysis of visual motion: A comparison of neuronal
  and psychophysical performance},\ }\href
  {https://doi.org/10.1523/JNEUROSCI.12-12-04745.1992} {\bibfield  {journal}
  {\bibinfo  {journal} {J. Neurosci.}\ }\textbf {\bibinfo {volume} {12}},\
  \bibinfo {pages} {4745} (\bibinfo {year} {1992})}\BibitemShut {NoStop}%
\bibitem [{\citenamefont {Wachowiak}\ and\ \citenamefont
  {Cohen}(2001)}]{wachowiak_representation_2001}%
  \BibitemOpen
  \bibfield  {author} {\bibinfo {author} {\bibfnamefont {M.}~\bibnamefont
  {Wachowiak}}\ and\ \bibinfo {author} {\bibfnamefont {L.~B.}\ \bibnamefont
  {Cohen}},\ }\bibfield  {title} {\bibinfo {title} {Representation of
  {{Odorants}} by {{Receptor Neuron Input}} to the {{Mouse Olfactory Bulb}}},\
  }\href {https://doi.org/10.1016/S0896-6273(01)00506-2} {\bibfield  {journal}
  {\bibinfo  {journal} {Neuron}\ }\textbf {\bibinfo {volume} {32}},\ \bibinfo
  {pages} {723} (\bibinfo {year} {2001})}\BibitemShut {NoStop}%
\bibitem [{\citenamefont {Evans}(1981)}]{evans_dynamic_1981}%
  \BibitemOpen
  \bibfield  {author} {\bibinfo {author} {\bibfnamefont {E.~F.}\ \bibnamefont
  {Evans}},\ }\bibfield  {title} {\bibinfo {title} {The {{Dynamic Range
  Problem}}: {{Place}} and {{Time Coding}} at the {{Level}} of {{Cochlear
  Nerve}} and {{Nucleus}}},\ }in\ \href
  {https://doi.org/10.1007/978-1-4684-3908-3_9} {\emph {\bibinfo {booktitle}
  {Neuronal {{Mechanisms}} of {{Hearing}}}}},\ \bibinfo {editor} {edited by\
  \bibinfo {editor} {\bibfnamefont {J.}~\bibnamefont {Syka}}\ and\ \bibinfo
  {editor} {\bibfnamefont {L.}~\bibnamefont {Aitkin}}}\ (\bibinfo  {publisher}
  {{Springer US}},\ \bibinfo {address} {{Boston, MA}},\ \bibinfo {year}
  {1981})\ pp.\ \bibinfo {pages} {69--85}\BibitemShut {NoStop}%
\bibitem [{\citenamefont {Dean}\ \emph {et~al.}(2005)\citenamefont {Dean},
  \citenamefont {Harper},\ and\ \citenamefont {McAlpine}}]{dean_neural_2005}%
  \BibitemOpen
  \bibfield  {author} {\bibinfo {author} {\bibfnamefont {I.}~\bibnamefont
  {Dean}}, \bibinfo {author} {\bibfnamefont {N.~S.}\ \bibnamefont {Harper}},\
  and\ \bibinfo {author} {\bibfnamefont {D.}~\bibnamefont {McAlpine}},\
  }\bibfield  {title} {\bibinfo {title} {Neural population coding of sound
  level adapts to stimulus statistics},\ }\href
  {https://doi.org/10.1038/nn1541} {\bibfield  {journal} {\bibinfo  {journal}
  {Nat. Neurosci.}\ }\textbf {\bibinfo {volume} {8}},\ \bibinfo {pages} {1684}
  (\bibinfo {year} {2005})}\BibitemShut {NoStop}%
\bibitem [{\citenamefont {Kinouchi}\ and\ \citenamefont
  {Copelli}(2006{\natexlab{a}})}]{kinouchi_optimal_2006}%
  \BibitemOpen
  \bibfield  {author} {\bibinfo {author} {\bibfnamefont {O.}~\bibnamefont
  {Kinouchi}}\ and\ \bibinfo {author} {\bibfnamefont {M.}~\bibnamefont
  {Copelli}},\ }\bibfield  {title} {\bibinfo {title} {Optimal dynamical range
  of excitable networks at criticality},\ }\href
  {https://doi.org/10.1038/nphys289} {\bibfield  {journal} {\bibinfo  {journal}
  {Nat. Phys.}\ }\textbf {\bibinfo {volume} {2}},\ \bibinfo {pages} {348}
  (\bibinfo {year} {2006}{\natexlab{a}})}\BibitemShut {NoStop}%
\bibitem [{\citenamefont {Gautam}\ \emph
  {et~al.}(2015{\natexlab{a}})\citenamefont {Gautam}, \citenamefont {Hoang},
  \citenamefont {McClanahan}, \citenamefont {Grady},\ and\ \citenamefont
  {Shew}}]{gautam2015maximizing}%
  \BibitemOpen
  \bibfield  {author} {\bibinfo {author} {\bibfnamefont {S.~H.}\ \bibnamefont
  {Gautam}}, \bibinfo {author} {\bibfnamefont {T.~T.}\ \bibnamefont {Hoang}},
  \bibinfo {author} {\bibfnamefont {K.}~\bibnamefont {McClanahan}}, \bibinfo
  {author} {\bibfnamefont {S.~K.}\ \bibnamefont {Grady}},\ and\ \bibinfo
  {author} {\bibfnamefont {W.~L.}\ \bibnamefont {Shew}},\ }\bibfield  {title}
  {\bibinfo {title} {Maximizing sensory dynamic range by tuning the cortical
  state to criticality},\ }\href@noop {} {\bibfield  {journal} {\bibinfo
  {journal} {PLoS computational biology}\ }\textbf {\bibinfo {volume} {11}},\
  \bibinfo {pages} {e1004576} (\bibinfo {year}
  {2015}{\natexlab{a}})}\BibitemShut {NoStop}%
\bibitem [{\citenamefont {Zierenberg}\ \emph
  {et~al.}(2020{\natexlab{a}})\citenamefont {Zierenberg}, \citenamefont
  {Wilting}, \citenamefont {Priesemann},\ and\ \citenamefont
  {Levina}}]{zierenberg_tailored_2020}%
  \BibitemOpen
  \bibfield  {author} {\bibinfo {author} {\bibfnamefont {J.}~\bibnamefont
  {Zierenberg}}, \bibinfo {author} {\bibfnamefont {J.}~\bibnamefont {Wilting}},
  \bibinfo {author} {\bibfnamefont {V.}~\bibnamefont {Priesemann}},\ and\
  \bibinfo {author} {\bibfnamefont {A.}~\bibnamefont {Levina}},\ }\bibfield
  {title} {\bibinfo {title} {Tailored ensembles of neural networks optimize
  sensitivity to stimulus statistics},\ }\href
  {https://doi.org/10.1103/PhysRevResearch.2.013115} {\bibfield  {journal}
  {\bibinfo  {journal} {Phys. Rev. Res.}\ }\textbf {\bibinfo {volume} {2}},\
  \bibinfo {pages} {013115} (\bibinfo {year} {2020}{\natexlab{a}})}\BibitemShut
  {NoStop}%
\bibitem [{\citenamefont {Hopfield}(1982)}]{hopfield_neural_1982}%
  \BibitemOpen
  \bibfield  {author} {\bibinfo {author} {\bibfnamefont {J.~J.}\ \bibnamefont
  {Hopfield}},\ }\bibfield  {title} {\bibinfo {title} {Neural networks and
  physical systems with emergent collective computational abilities},\ }\href
  {https://doi.org/10.1073/pnas.79.8.2554} {\bibfield  {journal} {\bibinfo
  {journal} {Proc. Natl. Acad. Sci.}\ }\textbf {\bibinfo {volume} {79}},\
  \bibinfo {pages} {2554} (\bibinfo {year} {1982})}\BibitemShut {NoStop}%
\bibitem [{\citenamefont {Battista}\ and\ \citenamefont
  {Monasson}(2020)}]{battista_capacity-resolution_2020}%
  \BibitemOpen
  \bibfield  {author} {\bibinfo {author} {\bibfnamefont {A.}~\bibnamefont
  {Battista}}\ and\ \bibinfo {author} {\bibfnamefont {R.}~\bibnamefont
  {Monasson}},\ }\bibfield  {title} {\bibinfo {title} {Capacity-{{Resolution
  Trade-Off}} in the {{Optimal Learning}} of {{Multiple Low-Dimensional
  Manifolds}} by {{Attractor Neural Networks}}},\ }\href@noop {} {\bibfield
  {journal} {\bibinfo  {journal} {Phys. Rev. Lett.}\ }\textbf {\bibinfo
  {volume} {124}},\ \bibinfo {pages} {5} (\bibinfo {year} {2020})}\BibitemShut
  {NoStop}%
\bibitem [{\citenamefont {Rabinovich}\ \emph {et~al.}(2008)\citenamefont
  {Rabinovich}, \citenamefont {Huerta},\ and\ \citenamefont
  {Laurent}}]{rabinovich_transient_2008}%
  \BibitemOpen
  \bibfield  {author} {\bibinfo {author} {\bibfnamefont {M.}~\bibnamefont
  {Rabinovich}}, \bibinfo {author} {\bibfnamefont {R.}~\bibnamefont {Huerta}},\
  and\ \bibinfo {author} {\bibfnamefont {G.}~\bibnamefont {Laurent}},\
  }\bibfield  {title} {\bibinfo {title} {Transient {{Dynamics}} for {{Neural
  Processing}}},\ }\href {https://doi.org/10.1126/science.1155564} {\bibfield
  {journal} {\bibinfo  {journal} {Science}\ }\textbf {\bibinfo {volume}
  {321}},\ \bibinfo {pages} {48} (\bibinfo {year} {2008})}\BibitemShut
  {NoStop}%
\bibitem [{\citenamefont {Gallego}\ \emph {et~al.}(2017)\citenamefont
  {Gallego}, \citenamefont {Perich}, \citenamefont {Miller},\ and\
  \citenamefont {Solla}}]{gallego_neural_2017}%
  \BibitemOpen
  \bibfield  {author} {\bibinfo {author} {\bibfnamefont {J.~A.}\ \bibnamefont
  {Gallego}}, \bibinfo {author} {\bibfnamefont {M.~G.}\ \bibnamefont {Perich}},
  \bibinfo {author} {\bibfnamefont {L.~E.}\ \bibnamefont {Miller}},\ and\
  \bibinfo {author} {\bibfnamefont {S.~A.}\ \bibnamefont {Solla}},\ }\bibfield
  {title} {\bibinfo {title} {Neural {{Manifolds}} for the {{Control}} of
  {{Movement}}},\ }\href {https://doi.org/10.1016/j.neuron.2017.05.025}
  {\bibfield  {journal} {\bibinfo  {journal} {Neuron}\ }\textbf {\bibinfo
  {volume} {94}},\ \bibinfo {pages} {978} (\bibinfo {year} {2017})}\BibitemShut
  {NoStop}%
\bibitem [{\citenamefont {Vyas}\ \emph {et~al.}(2020)\citenamefont {Vyas},
  \citenamefont {Golub}, \citenamefont {Sussillo},\ and\ \citenamefont
  {Shenoy}}]{vyas_computation_2020}%
  \BibitemOpen
  \bibfield  {author} {\bibinfo {author} {\bibfnamefont {S.}~\bibnamefont
  {Vyas}}, \bibinfo {author} {\bibfnamefont {M.~D.}\ \bibnamefont {Golub}},
  \bibinfo {author} {\bibfnamefont {D.}~\bibnamefont {Sussillo}},\ and\
  \bibinfo {author} {\bibfnamefont {K.~V.}\ \bibnamefont {Shenoy}},\ }\bibfield
   {title} {\bibinfo {title} {Computation {{Through Neural Population
  Dynamics}}},\ }\href {https://doi.org/10.1146/annurev-neuro-092619-094115}
  {\bibfield  {journal} {\bibinfo  {journal} {Annu. Rev. Neurosci.}\ }\textbf
  {\bibinfo {volume} {43}},\ \bibinfo {pages} {249} (\bibinfo {year}
  {2020})}\BibitemShut {NoStop}%
\bibitem [{\citenamefont {Beggs}(2008)}]{beggs_criticality_2008}%
  \BibitemOpen
  \bibfield  {author} {\bibinfo {author} {\bibfnamefont {J.~M.}\ \bibnamefont
  {Beggs}},\ }\bibfield  {title} {\bibinfo {title} {The criticality hypothesis:
  How local cortical networks might optimize information processing},\ }\href
  {https://doi.org/10.1098/rsta.2007.2092} {\bibfield  {journal} {\bibinfo
  {journal} {Philos. Trans. R. Soc. Lond. Math. Phys. Eng. Sci.}\ }\textbf
  {\bibinfo {volume} {366}},\ \bibinfo {pages} {329} (\bibinfo {year}
  {2008})}\BibitemShut {NoStop}%
\bibitem [{\citenamefont {Chialvo}(2010)}]{chialvo_emergent_2010}%
  \BibitemOpen
  \bibfield  {author} {\bibinfo {author} {\bibfnamefont {D.~R.}\ \bibnamefont
  {Chialvo}},\ }\bibfield  {title} {\bibinfo {title} {Emergent complex neural
  dynamics},\ }\href {https://doi.org/10.1038/nphys1803} {\bibfield  {journal}
  {\bibinfo  {journal} {Nat. Phys.}\ }\textbf {\bibinfo {volume} {6}},\
  \bibinfo {pages} {744} (\bibinfo {year} {2010})}\BibitemShut {NoStop}%
\bibitem [{\citenamefont {Mu{\~n}oz}(2018)}]{munoz_colloquium_2018}%
  \BibitemOpen
  \bibfield  {author} {\bibinfo {author} {\bibfnamefont {M.~A.}\ \bibnamefont
  {Mu{\~n}oz}},\ }\bibfield  {title} {\bibinfo {title} {Colloquium:
  {{Criticality}} and dynamical scaling in living systems},\ }\href
  {https://doi.org/10.1103/RevModPhys.90.031001} {\bibfield  {journal}
  {\bibinfo  {journal} {Rev. Mod. Phys.}\ }\textbf {\bibinfo {volume} {90}},\
  \bibinfo {pages} {031001} (\bibinfo {year} {2018})}\BibitemShut {NoStop}%
\bibitem [{\citenamefont {Wilting}\ \emph {et~al.}(2018)\citenamefont
  {Wilting}, \citenamefont {Dehning}, \citenamefont {Pinheiro~Neto},
  \citenamefont {Rudelt}, \citenamefont {Wibral}, \citenamefont {Zierenberg},\
  and\ \citenamefont {Priesemann}}]{wilting_operating_2018}%
  \BibitemOpen
  \bibfield  {author} {\bibinfo {author} {\bibfnamefont {J.}~\bibnamefont
  {Wilting}}, \bibinfo {author} {\bibfnamefont {J.}~\bibnamefont {Dehning}},
  \bibinfo {author} {\bibfnamefont {J.}~\bibnamefont {Pinheiro~Neto}}, \bibinfo
  {author} {\bibfnamefont {L.}~\bibnamefont {Rudelt}}, \bibinfo {author}
  {\bibfnamefont {M.}~\bibnamefont {Wibral}}, \bibinfo {author} {\bibfnamefont
  {J.}~\bibnamefont {Zierenberg}},\ and\ \bibinfo {author} {\bibfnamefont
  {V.}~\bibnamefont {Priesemann}},\ }\bibfield  {title} {\bibinfo {title}
  {Operating in a {{Reverberating Regime Enables Rapid Tuning}} of {{Network
  States}} to {{Task Requirements}}},\ }\href
  {https://doi.org/10.3389/fnsys.2018.00055} {\bibfield  {journal} {\bibinfo
  {journal} {Front. Syst. Neurosci.}\ }\textbf {\bibinfo {volume} {12}},\
  \bibinfo {pages} {55} (\bibinfo {year} {2018})}\BibitemShut {NoStop}%
\bibitem [{\citenamefont {Wilting}\ and\ \citenamefont
  {Priesemann}(2019)}]{wilting_25_2019}%
  \BibitemOpen
  \bibfield  {author} {\bibinfo {author} {\bibfnamefont {J.}~\bibnamefont
  {Wilting}}\ and\ \bibinfo {author} {\bibfnamefont {V.}~\bibnamefont
  {Priesemann}},\ }\bibfield  {title} {\bibinfo {title} {25 years of
  criticality in neuroscience \textemdash{} established results, open
  controversies, novel concepts},\ }\href
  {https://doi.org/10.1016/j.conb.2019.08.002} {\bibfield  {journal} {\bibinfo
  {journal} {Curr. Opin. Neurobiol.}\ }\bibinfo {series} {Computational
  {{Neuroscience}}},\ \textbf {\bibinfo {volume} {58}},\ \bibinfo {pages} {105}
  (\bibinfo {year} {2019})}\BibitemShut {NoStop}%
\bibitem [{\citenamefont {Larremore}\ \emph {et~al.}(2011)\citenamefont
  {Larremore}, \citenamefont {Shew},\ and\ \citenamefont
  {Restrepo}}]{larremore2011predicting}%
  \BibitemOpen
  \bibfield  {author} {\bibinfo {author} {\bibfnamefont {D.~B.}\ \bibnamefont
  {Larremore}}, \bibinfo {author} {\bibfnamefont {W.~L.}\ \bibnamefont
  {Shew}},\ and\ \bibinfo {author} {\bibfnamefont {J.~G.}\ \bibnamefont
  {Restrepo}},\ }\bibfield  {title} {\bibinfo {title} {Predicting criticality
  and dynamic range in complex networks: effects of topology},\ }\href@noop {}
  {\bibfield  {journal} {\bibinfo  {journal} {Physical review letters}\
  }\textbf {\bibinfo {volume} {106}},\ \bibinfo {pages} {058101} (\bibinfo
  {year} {2011})}\BibitemShut {NoStop}%
\bibitem [{\citenamefont {Berens}\ \emph
  {et~al.}(2009{\natexlab{a}})\citenamefont {Berens}, \citenamefont {Gerwinn},
  \citenamefont {Ecker},\ and\ \citenamefont
  {Bethge}}]{berens_neurometric_2009}%
  \BibitemOpen
  \bibfield  {author} {\bibinfo {author} {\bibfnamefont {P.}~\bibnamefont
  {Berens}}, \bibinfo {author} {\bibfnamefont {S.}~\bibnamefont {Gerwinn}},
  \bibinfo {author} {\bibfnamefont {A.}~\bibnamefont {Ecker}},\ and\ \bibinfo
  {author} {\bibfnamefont {M.}~\bibnamefont {Bethge}},\ }\bibfield  {title}
  {\bibinfo {title} {Neurometric function analysis of population codes},\ }in\
  \href@noop {} {\emph {\bibinfo {booktitle} {Adv. {{Neural Inf}}. {{Process}}.
  {{Syst}}.}}},\ Vol.~\bibinfo {volume} {22},\ \bibinfo {editor} {edited by\
  \bibinfo {editor} {\bibfnamefont {Y.}~\bibnamefont {Bengio}}, \bibinfo
  {editor} {\bibfnamefont {D.}~\bibnamefont {Schuurmans}}, \bibinfo {editor}
  {\bibfnamefont {J.}~\bibnamefont {Lafferty}}, \bibinfo {editor}
  {\bibfnamefont {C.}~\bibnamefont {Williams}},\ and\ \bibinfo {editor}
  {\bibfnamefont {A.}~\bibnamefont {Culotta}}}\ (\bibinfo  {publisher} {{Curran
  Associates, Inc.}},\ \bibinfo {year} {2009})\BibitemShut {NoStop}%
\bibitem [{\citenamefont {Kinouchi}\ and\ \citenamefont
  {Copelli}(2006{\natexlab{b}})}]{kinouchi2006optimal}%
  \BibitemOpen
  \bibfield  {author} {\bibinfo {author} {\bibfnamefont {O.}~\bibnamefont
  {Kinouchi}}\ and\ \bibinfo {author} {\bibfnamefont {M.}~\bibnamefont
  {Copelli}},\ }\bibfield  {title} {\bibinfo {title} {Optimal dynamical range
  of excitable networks at criticality},\ }\href@noop {} {\bibfield  {journal}
  {\bibinfo  {journal} {Nature physics}\ }\textbf {\bibinfo {volume} {2}},\
  \bibinfo {pages} {348} (\bibinfo {year} {2006}{\natexlab{b}})}\BibitemShut
  {NoStop}%
\bibitem [{\citenamefont {Shew}\ \emph {et~al.}(2009)\citenamefont {Shew},
  \citenamefont {Yang}, \citenamefont {Petermann}, \citenamefont {Roy},\ and\
  \citenamefont {Plenz}}]{shew_neuronal_2009}%
  \BibitemOpen
  \bibfield  {author} {\bibinfo {author} {\bibfnamefont {W.~L.}\ \bibnamefont
  {Shew}}, \bibinfo {author} {\bibfnamefont {H.}~\bibnamefont {Yang}}, \bibinfo
  {author} {\bibfnamefont {T.}~\bibnamefont {Petermann}}, \bibinfo {author}
  {\bibfnamefont {R.}~\bibnamefont {Roy}},\ and\ \bibinfo {author}
  {\bibfnamefont {D.}~\bibnamefont {Plenz}},\ }\bibfield  {title} {\bibinfo
  {title} {Neuronal {{Avalanches Imply Maximum Dynamic Range}} in {{Cortical
  Networks}} at {{Criticality}}},\ }\href
  {https://doi.org/10.1523/JNEUROSCI.3864-09.2009} {\bibfield  {journal}
  {\bibinfo  {journal} {J. Neurosci.}\ }\textbf {\bibinfo {volume} {29}},\
  \bibinfo {pages} {15595} (\bibinfo {year} {2009})}\BibitemShut {NoStop}%
\bibitem [{\citenamefont {{Shwartz-Ziv}}\ and\ \citenamefont
  {Tishby}(2017)}]{shwartz-ziv_opening_2017}%
  \BibitemOpen
  \bibfield  {author} {\bibinfo {author} {\bibfnamefont {R.}~\bibnamefont
  {{Shwartz-Ziv}}}\ and\ \bibinfo {author} {\bibfnamefont {N.}~\bibnamefont
  {Tishby}},\ }\bibfield  {title} {\bibinfo {title} {Opening the {{Black Box}}
  of {{Deep Neural Networks}} via {{Information}}},\ }\href@noop {} {\bibfield
  {journal} {\bibinfo  {journal} {ArXiv170300810 Cs}\ } (\bibinfo {year}
  {2017})},\ \Eprint {https://arxiv.org/abs/1703.00810} {arxiv:1703.00810 [cs]}
  \BibitemShut {NoStop}%
\bibitem [{\citenamefont {Shew}\ and\ \citenamefont
  {Plenz}(2013)}]{shew_functional_2013}%
  \BibitemOpen
  \bibfield  {author} {\bibinfo {author} {\bibfnamefont {W.~L.}\ \bibnamefont
  {Shew}}\ and\ \bibinfo {author} {\bibfnamefont {D.}~\bibnamefont {Plenz}},\
  }\bibfield  {title} {\bibinfo {title} {The {{Functional Benefits}} of
  {{Criticality}} in the {{Cortex}} , {{The Functional Benefits}} of
  {{Criticality}} in the {{Cortex}}},\ }\href
  {https://doi.org/10.1177/1073858412445487} {\bibfield  {journal} {\bibinfo
  {journal} {The Neuroscientist}\ }\textbf {\bibinfo {volume} {19}},\ \bibinfo
  {pages} {88} (\bibinfo {year} {2013})}\BibitemShut {NoStop}%
\bibitem [{\citenamefont {Gautam}\ \emph
  {et~al.}(2015{\natexlab{b}})\citenamefont {Gautam}, \citenamefont {Hoang},
  \citenamefont {McClanahan}, \citenamefont {Grady},\ and\ \citenamefont
  {Shew}}]{gautam_maximizing_2015}%
  \BibitemOpen
  \bibfield  {author} {\bibinfo {author} {\bibfnamefont {S.~H.}\ \bibnamefont
  {Gautam}}, \bibinfo {author} {\bibfnamefont {T.~T.}\ \bibnamefont {Hoang}},
  \bibinfo {author} {\bibfnamefont {K.}~\bibnamefont {McClanahan}}, \bibinfo
  {author} {\bibfnamefont {S.~K.}\ \bibnamefont {Grady}},\ and\ \bibinfo
  {author} {\bibfnamefont {W.~L.}\ \bibnamefont {Shew}},\ }\bibfield  {title}
  {\bibinfo {title} {Maximizing {{Sensory Dynamic Range}} by {{Tuning}} the
  {{Cortical State}} to {{Criticality}}},\ }\href
  {https://doi.org/10.1371/journal.pcbi.1004576} {\bibfield  {journal}
  {\bibinfo  {journal} {PLoS Comput. Biol.}\ }\textbf {\bibinfo {volume}
  {11}},\ \bibinfo {pages} {e1004576} (\bibinfo {year}
  {2015}{\natexlab{b}})}\BibitemShut {NoStop}%
\bibitem [{\citenamefont {Dahmen}\ \emph {et~al.}(2022)\citenamefont {Dahmen},
  \citenamefont {Recanatesi}, \citenamefont {Jia}, \citenamefont {Ocker},
  \citenamefont {Campagnola}, \citenamefont {Jarsky}, \citenamefont {Seeman},
  \citenamefont {Helias},\ and\ \citenamefont
  {{Shea-Brown}}}]{dahmen_strong_2022}%
  \BibitemOpen
  \bibfield  {author} {\bibinfo {author} {\bibfnamefont {D.}~\bibnamefont
  {Dahmen}}, \bibinfo {author} {\bibfnamefont {S.}~\bibnamefont {Recanatesi}},
  \bibinfo {author} {\bibfnamefont {X.}~\bibnamefont {Jia}}, \bibinfo {author}
  {\bibfnamefont {G.~K.}\ \bibnamefont {Ocker}}, \bibinfo {author}
  {\bibfnamefont {L.}~\bibnamefont {Campagnola}}, \bibinfo {author}
  {\bibfnamefont {T.}~\bibnamefont {Jarsky}}, \bibinfo {author} {\bibfnamefont
  {S.}~\bibnamefont {Seeman}}, \bibinfo {author} {\bibfnamefont
  {M.}~\bibnamefont {Helias}},\ and\ \bibinfo {author} {\bibfnamefont
  {E.}~\bibnamefont {{Shea-Brown}}},\ }\bibfield  {title} {\bibinfo {title}
  {Strong and localized recurrence controls dimensionality of neural activity
  across brain areas},\ }\href {https://doi.org/10.1101/2020.11.02.365072}
  {\bibfield  {journal} {\bibinfo  {journal} {bioRxiv}\ ,\ \bibinfo {pages}
  {2020.11.02.365072}} (\bibinfo {year} {2022})}\BibitemShut {NoStop}%
\bibitem [{\citenamefont {Zeraati}\ \emph {et~al.}(2023)\citenamefont
  {Zeraati}, \citenamefont {Shi}, \citenamefont {Steinmetz}, \citenamefont
  {Gieselmann}, \citenamefont {Thiele}, \citenamefont {Moore}, \citenamefont
  {Levina},\ and\ \citenamefont {Engel}}]{zeraati2023intrinsic}%
  \BibitemOpen
  \bibfield  {author} {\bibinfo {author} {\bibfnamefont {R.}~\bibnamefont
  {Zeraati}}, \bibinfo {author} {\bibfnamefont {Y.-L.}\ \bibnamefont {Shi}},
  \bibinfo {author} {\bibfnamefont {N.}~\bibnamefont {Steinmetz}}, \bibinfo
  {author} {\bibfnamefont {M.}~\bibnamefont {Gieselmann}}, \bibinfo {author}
  {\bibfnamefont {A.}~\bibnamefont {Thiele}}, \bibinfo {author} {\bibfnamefont
  {T.}~\bibnamefont {Moore}}, \bibinfo {author} {\bibfnamefont
  {A.}~\bibnamefont {Levina}},\ and\ \bibinfo {author} {\bibfnamefont
  {T.}~\bibnamefont {Engel}},\ }\bibfield  {title} {\bibinfo {title} {Intrinsic
  timescales in the visual cortex change with selective attention and reflect
  spatial connectivity},\ }\href {https://doi.org/10.1038/s41467-023-37613-7}
  {\bibfield  {journal} {\bibinfo  {journal} {Nature Communications}\ }\textbf
  {\bibinfo {volume} {14}} (\bibinfo {year} {2023})}\BibitemShut {NoStop}%
\bibitem [{\citenamefont {Khajehabdollahi}\ \emph {et~al.}(2022)\citenamefont
  {Khajehabdollahi}, \citenamefont {Prosi}, \citenamefont {Giannakakis},
  \citenamefont {Martius},\ and\ \citenamefont
  {Levina}}]{khajehabdollahi2022critical}%
  \BibitemOpen
  \bibfield  {author} {\bibinfo {author} {\bibfnamefont {S.}~\bibnamefont
  {Khajehabdollahi}}, \bibinfo {author} {\bibfnamefont {J.}~\bibnamefont
  {Prosi}}, \bibinfo {author} {\bibfnamefont {E.}~\bibnamefont {Giannakakis}},
  \bibinfo {author} {\bibfnamefont {G.}~\bibnamefont {Martius}},\ and\ \bibinfo
  {author} {\bibfnamefont {A.}~\bibnamefont {Levina}},\ }\bibfield  {title}
  {\bibinfo {title} {When to be critical? performance and evolvability in
  different regimes of neural ising agents},\ }\href@noop {} {\bibfield
  {journal} {\bibinfo  {journal} {Artificial Life}\ }\textbf {\bibinfo {volume}
  {28}},\ \bibinfo {pages} {458} (\bibinfo {year} {2022})}\BibitemShut
  {NoStop}%
\bibitem [{\citenamefont {Lindner}\ \emph {et~al.}(2011)\citenamefont
  {Lindner}, \citenamefont {Vicente}, \citenamefont {Priesemann},\ and\
  \citenamefont {Wibral}}]{lindner_trentool_2011}%
  \BibitemOpen
  \bibfield  {author} {\bibinfo {author} {\bibfnamefont {M.}~\bibnamefont
  {Lindner}}, \bibinfo {author} {\bibfnamefont {R.}~\bibnamefont {Vicente}},
  \bibinfo {author} {\bibfnamefont {V.}~\bibnamefont {Priesemann}},\ and\
  \bibinfo {author} {\bibfnamefont {M.}~\bibnamefont {Wibral}},\ }\bibfield
  {title} {\bibinfo {title} {{{TRENTOOL}}: {{A Matlab}} open source toolbox to
  analyse information flow in time series data with transfer entropy},\ }\href
  {https://doi.org/10.1186/1471-2202-12-119} {\bibfield  {journal} {\bibinfo
  {journal} {BMC Neurosci.}\ }\textbf {\bibinfo {volume} {12}},\ \bibinfo
  {pages} {119} (\bibinfo {year} {2011})}\BibitemShut {NoStop}%
\bibitem [{\citenamefont {Wollstadt}\ \emph {et~al.}(2019)\citenamefont
  {Wollstadt}, \citenamefont {Lizier}, \citenamefont {Vicente}, \citenamefont
  {Finn}, \citenamefont {{Martinez-Zarzuela}}, \citenamefont {Mediano},
  \citenamefont {Novelli},\ and\ \citenamefont
  {Wibral}}]{wollstadt_idtxl_2019}%
  \BibitemOpen
  \bibfield  {author} {\bibinfo {author} {\bibfnamefont {P.}~\bibnamefont
  {Wollstadt}}, \bibinfo {author} {\bibfnamefont {J.~T.}\ \bibnamefont
  {Lizier}}, \bibinfo {author} {\bibfnamefont {R.}~\bibnamefont {Vicente}},
  \bibinfo {author} {\bibfnamefont {C.}~\bibnamefont {Finn}}, \bibinfo {author}
  {\bibfnamefont {M.}~\bibnamefont {{Martinez-Zarzuela}}}, \bibinfo {author}
  {\bibfnamefont {P.}~\bibnamefont {Mediano}}, \bibinfo {author} {\bibfnamefont
  {L.}~\bibnamefont {Novelli}},\ and\ \bibinfo {author} {\bibfnamefont
  {M.}~\bibnamefont {Wibral}},\ }\bibfield  {title} {\bibinfo {title}
  {{{IDTxl}}: {{The Information Dynamics Toolkit}} xl: A {{Python}} package for
  the efficient analysis of multivariate information dynamics in networks},\
  }\href {https://doi.org/10.21105/joss.01081} {\bibfield  {journal} {\bibinfo
  {journal} {J. Open Source Softw.}\ }\textbf {\bibinfo {volume} {4}},\
  \bibinfo {pages} {1081} (\bibinfo {year} {2019})}\BibitemShut {NoStop}%
\bibitem [{\citenamefont {Felleman}\ and\ \citenamefont
  {Van}(1991)}]{felleman_distributed_1991}%
  \BibitemOpen
  \bibfield  {author} {\bibinfo {author} {\bibfnamefont {D.~J.}\ \bibnamefont
  {Felleman}}\ and\ \bibinfo {author} {\bibfnamefont {D.~E.}\ \bibnamefont
  {Van}},\ }\bibfield  {title} {\bibinfo {title} {Distributed hierarchical
  processing in the primate cerebral cortex},\ }\href
  {https://doi.org/10.1093/cercor/1.1.1} {\bibfield  {journal} {\bibinfo
  {journal} {Cereb. Cortex}\ }\textbf {\bibinfo {volume} {1}},\ \bibinfo
  {pages} {1} (\bibinfo {year} {1991})}\BibitemShut {NoStop}%
\bibitem [{\citenamefont {Cramer}\ \emph {et~al.}(2020)\citenamefont {Cramer},
  \citenamefont {St{\"o}ckel}, \citenamefont {Kreft}, \citenamefont {Wibral},
  \citenamefont {Schemmel}, \citenamefont {Meier},\ and\ \citenamefont
  {Priesemann}}]{cramer_control_2020}%
  \BibitemOpen
  \bibfield  {author} {\bibinfo {author} {\bibfnamefont {B.}~\bibnamefont
  {Cramer}}, \bibinfo {author} {\bibfnamefont {D.}~\bibnamefont {St{\"o}ckel}},
  \bibinfo {author} {\bibfnamefont {M.}~\bibnamefont {Kreft}}, \bibinfo
  {author} {\bibfnamefont {M.}~\bibnamefont {Wibral}}, \bibinfo {author}
  {\bibfnamefont {J.}~\bibnamefont {Schemmel}}, \bibinfo {author}
  {\bibfnamefont {K.}~\bibnamefont {Meier}},\ and\ \bibinfo {author}
  {\bibfnamefont {V.}~\bibnamefont {Priesemann}},\ }\bibfield  {title}
  {\bibinfo {title} {Control of criticality and computation in spiking
  neuromorphic networks with plasticity},\ }\href
  {https://doi.org/10.1038/s41467-020-16548-3} {\bibfield  {journal} {\bibinfo
  {journal} {Nat. Commun.}\ }\textbf {\bibinfo {volume} {11}},\ \bibinfo
  {pages} {2853} (\bibinfo {year} {2020})}\BibitemShut {NoStop}%
\bibitem [{\citenamefont {Buend{\'i}a}\ \emph {et~al.}(2019)\citenamefont
  {Buend{\'i}a}, \citenamefont {Villegas}, \citenamefont {di~Santo},
  \citenamefont {Vezzani}, \citenamefont {Burioni},\ and\ \citenamefont
  {Mu{\~n}oz}}]{buendia_jensens_2019}%
  \BibitemOpen
  \bibfield  {author} {\bibinfo {author} {\bibfnamefont {V.}~\bibnamefont
  {Buend{\'i}a}}, \bibinfo {author} {\bibfnamefont {P.}~\bibnamefont
  {Villegas}}, \bibinfo {author} {\bibfnamefont {S.}~\bibnamefont {di~Santo}},
  \bibinfo {author} {\bibfnamefont {A.}~\bibnamefont {Vezzani}}, \bibinfo
  {author} {\bibfnamefont {R.}~\bibnamefont {Burioni}},\ and\ \bibinfo {author}
  {\bibfnamefont {M.~A.}\ \bibnamefont {Mu{\~n}oz}},\ }\bibfield  {title}
  {\bibinfo {title} {Jensen's force and the statistical mechanics of cortical
  asynchronous states},\ }\href {https://doi.org/10.1038/s41598-019-51520-2}
  {\bibfield  {journal} {\bibinfo  {journal} {Sci. Rep.}\ }\textbf {\bibinfo
  {volume} {9}},\ \bibinfo {pages} {1} (\bibinfo {year} {2019})}\BibitemShut
  {NoStop}%
\bibitem [{\citenamefont {Zierenberg}\ \emph
  {et~al.}(2020{\natexlab{b}})\citenamefont {Zierenberg}, \citenamefont
  {Wilting}, \citenamefont {Priesemann},\ and\ \citenamefont
  {Levina}}]{zierenberg_description_2020}%
  \BibitemOpen
  \bibfield  {author} {\bibinfo {author} {\bibfnamefont {J.}~\bibnamefont
  {Zierenberg}}, \bibinfo {author} {\bibfnamefont {J.}~\bibnamefont {Wilting}},
  \bibinfo {author} {\bibfnamefont {V.}~\bibnamefont {Priesemann}},\ and\
  \bibinfo {author} {\bibfnamefont {A.}~\bibnamefont {Levina}},\ }\bibfield
  {title} {\bibinfo {title} {Description of spreading dynamics by microscopic
  network models and macroscopic branching processes can differ due to
  coalescence},\ }\href {https://doi.org/10.1103/PhysRevE.101.022301}
  {\bibfield  {journal} {\bibinfo  {journal} {Phys. Rev. E}\ }\textbf {\bibinfo
  {volume} {101}},\ \bibinfo {pages} {022301} (\bibinfo {year}
  {2020}{\natexlab{b}})}\BibitemShut {NoStop}%
\bibitem [{\citenamefont {Henkel}\ \emph {et~al.}(2008)\citenamefont {Henkel},
  \citenamefont {Hinrichsen},\ and\ \citenamefont
  {L{\"u}beck}}]{henkel_non-equilibrium_2008}%
  \BibitemOpen
  \bibfield  {author} {\bibinfo {author} {\bibfnamefont {M.}~\bibnamefont
  {Henkel}}, \bibinfo {author} {\bibfnamefont {H.}~\bibnamefont {Hinrichsen}},\
  and\ \bibinfo {author} {\bibfnamefont {S.}~\bibnamefont {L{\"u}beck}},\
  }\href@noop {} {\emph {\bibinfo {title} {Non-{{Equilibrium Phase Transitions
  Volume}} 1: {{Absorbing Phase Transitions}}}}}\ (\bibinfo  {publisher}
  {{Springer Science \& Business Media}},\ \bibinfo {year} {2008})\BibitemShut
  {NoStop}%
\bibitem [{\citenamefont {Risken}\ and\ \citenamefont
  {Frank}(2012)}]{risken_fokker-planck_2012}%
  \BibitemOpen
  \bibfield  {author} {\bibinfo {author} {\bibfnamefont {H.}~\bibnamefont
  {Risken}}\ and\ \bibinfo {author} {\bibfnamefont {T.}~\bibnamefont {Frank}},\
  }\href@noop {} {\emph {\bibinfo {title} {The {{Fokker-Planck Equation}}:
  {{Methods}} of {{Solution}} and {{Applications}}}}}\ (\bibinfo  {publisher}
  {{Springer Science \& Business Media}},\ \bibinfo {year} {2012})\BibitemShut
  {NoStop}%
\bibitem [{\citenamefont {Berens}\ \emph
  {et~al.}(2009{\natexlab{b}})\citenamefont {Berens}, \citenamefont {Gerwinn},
  \citenamefont {Ecker},\ and\ \citenamefont {Bethge}}]{berens2009neurometric}%
  \BibitemOpen
  \bibfield  {author} {\bibinfo {author} {\bibfnamefont {P.}~\bibnamefont
  {Berens}}, \bibinfo {author} {\bibfnamefont {S.}~\bibnamefont {Gerwinn}},
  \bibinfo {author} {\bibfnamefont {A.~S.}\ \bibnamefont {Ecker}},\ and\
  \bibinfo {author} {\bibfnamefont {M.}~\bibnamefont {Bethge}},\ }\bibfield
  {title} {\bibinfo {title} {Neurometric function analysis of population
  codes},\ }\href@noop {} {\bibfield  {journal} {\bibinfo  {journal} {IJS
  ($\theta$1, $\theta$2)}\ }\textbf {\bibinfo {volume} {1}},\ \bibinfo {pages}
  {2} (\bibinfo {year} {2009}{\natexlab{b}})}\BibitemShut {NoStop}%
\end{thebibliography}%

\clearpage
\begingroup
\section*{Methods}
\paragraph{Neural Network Model:} 
We consider a network of $N$ binary spiking neurons. 
Each neuron is described by a state variable $s_i$ and can be active ($s_i(t)=1$) or inactive ($s_i(t)=0$)). 
The total network activity is described by the vector $s(t)= (s_1(t), \ldots , s_N(t)$ where $t$ evolves in discrete steps of $\Delta t=\SI{1}{ms}$.
Neurons can be activated by recurrent input from other neurons via a probabilistic integrate-and-fire mechanism with probability $p^\mathrm{rec}[s_i(t+\Delta t)=1 | s(t)]=f(\sum_j w_{ij} s_j(t))$, where $w_{ij}$ are coupling weights (not symmetric) and $f(x)$ is a rectified linear function with $f(x)=0$ for $x<0$, $f(x)=x$ for $0<x<1$, and $f(x)=1$ for $x>1$.
We construct connectivity matrix $W = \left (w_{ij} \right)$ as a sparse matrix with $K$ randomly selected non-zero elements per row with value $w=\lambda/K$, such that the network has a fixed in-degree $K$.
This ensures that the largest eigenvalue is always $\lambda$ and thereby simplifies our analytic approach~\cite{buendia_jensens_2019}.
However, random connectivity with the same mean degree and mean incoming weighted strength ($\sum_j {w_{ij}} = \lambda$) show the same results due to self-averaging properties of random networks (\ref{supp3}).

From the $N$ neurons, a subset $N^\mathrm{in} = \mu N $ neurons receive external input in the form of Poisson noise with rate $h$ that activates each of these neurons with probability $p^\mathrm{ext}[s_{t+1}^i=1]=1-e^{-h\Delta t}$. 
The output is either recorded from $N^\mathrm{out}=(1-\mu)N$ randomly chosen neurons or from the full network. 
We call the \emph{input activity} the sum of states of the $N^\mathrm{in}$ input neurons, $A^\mathrm{in}_t=\sum_{i=1}^{N^\mathrm{in}} s^{\mathrm{in}}_i$, and the \emph{output activity}  the sum of the states of  $N^\mathrm{out}$ randomly output neurons as $A^\mathrm{out}_t=\sum_{i=1}^{N^\mathrm{out}} s^{\mathrm{out},i}_t$. 
The average activity over all neurons in each population is $a^\mathrm{\ast}_t=A^\mathrm{\ast}_t/N^\mathrm{\ast}$.

\paragraph{Continuous approximation of output distribution:}
The stochastic simulations provide an approximation for the probabilistic input-output relationship of the network, $P(A^\mathrm{out}|h_i)$. 
The output activity can assume only discrete values $A^\mathrm{out}\in[0,N^\mathrm{out}]$, resulting in discrete values of $a^\mathrm{out} \in [0,1]$. 
We obtain simulation results for a finite number of inputs $h_i$.
For each of them, we use a maximum likelihood to find parameters $\alpha$ and $\beta$ of a Beta distribution $\text{Beta}(\alpha,\beta)$ that best describes the observations.
The ML fits closely follow the empirical density, justifying the selection of Beta distribution (Fig.~\ref{fig:extData_workflow}A).
Our subsequent analysis requires estimates of output distribution for any possible input value, hence we construct a continuous approximation for many sampled inputs and interpolate for inputs between them.
Specifically, 
to obtain estimates for non-sampled inputs, we linearly interpolate the parameters $\alpha$ and $\beta$ between closest inputs(Fig.~\ref{fig:extData_workflow}B).
This way we can obtain approximate conditional density $p(o|h)$ for any $o \in [0,1]$ and $h \in [10^{-7},10^{1.5}]$.

\paragraph*{Mean-field solution for the limit \mbox{$T\to\infty$}:}
For infinite observation time, we can neglect fluctuations and the distribution of the output response for given input becomes a delta function, so each input correspond to single output. 
To find this output response, we separately compute activation of $N^\mathrm{in}=\mu N$ input neurons  $a^\mathrm{in}$ and the $N-N^\mathrm{in}=(1-\mu)N$ other, non-driven neurons $a^\mathrm{rest}$.
The mean probability to activate any neuron by recurrent inputs, $\overline{p^\mathrm{rec}}=\overline{\sum_{ij} w_{ij}s_j(t)}$, can be split into the contributions from the input and non-input populations as: 
\begin{equation}\label{eq_mean-field_prob}
    \overline{p^\mathrm{rec}}\approx\lambda\left[\mu a^\mathrm{in} +(1-\mu)a^\mathrm{rest})\right].
\end{equation}
Here we used that $\overline{\sum_i w_{ij}} = \overline{\sum_j w_{ij}} = \lambda$
In addition, input neurons can be independently activated by the external input with probability $p^\mathrm{ext}=1-e^{-h\Delta t}$, resulting in a total probability of their activation after taking into account coalescence~\cite{zierenberg_description_2020} being
\begin{equation*}
    \overline{p}=1-(1-\overline{p^\mathrm{rec}})(1-p^\mathrm{ext})
\end{equation*}
Since the average probability of activation is itself an approximation of the mean activity, we obtain the following system of self-consistent equations
\begin{align}\label{eq_mean-field_equation}
    a^\mathrm{in} &= 1-\left(1-\lambda\left[\mu a^\mathrm{in} + (1-\mu)a^\mathrm{rest}\right]\right)e^{-h\Delta t}\nonumber\\
    a^\mathrm{rest} &=\lambda\left[\mu a^\mathrm{in} + (1-\mu)a^\mathrm{rest}\right]
\end{align} 
which are solved by
\begin{align}
    a^\mathrm{in}&=\frac{\left(1-e^{-h\Delta t}\right)\left(1-\lambda\left[1-\mu\right]\right)}{1-\lambda\left(1-\mu\right)-\lambda \mu e^{-h\Delta t}}\label{eq_mean-field_equation_in}\\
    a^\mathrm{rest}&=\frac{\lambda \mu (1-e^{-h\Delta t})}{1-\lambda\left(1-\mu\right)-\lambda \mu e^{-h\Delta t}}\label{eq_mean-field_equation_rest}
\end{align}
Finally, we obtain an estimate of the mean-field output response by a combination of these results. 
If the output neurons span the full population or are randomly chosen from the full population, then the mean-field response is 
\begin{align}\label{eq:mean}
    a^\mathrm{out}&=\mu a^\mathrm{in}+\left(1-\mu\right) a^\mathrm{rest}\\
    &=\frac{\mu (1-e^{-h\Delta t})}{1-\lambda\left(1-\mu\right)-\lambda \mu e^{-h\Delta t}}\nonumber
\end{align}

\paragraph*{Analytic solution for the limit $T\to 0$:}
At the level of instantaneous activity, we can derive an approximate analytic solution for the distribution of network activity by solving the corresponding Fokker-Planck equation. 
To arrive at the Fokker-Plank equation, we start from a master equation where the activation and deactivation of neurons are treated as a birth-death process. 
Technically this assumes asynchronous updates of the network state, i.e., sequentially updating randomly selected neurons, which should relax to the same stationary activity distribution as synchronous updates employed in our simulations~\cite{henkel_non-equilibrium_2008}. 
The following considerations build on the mean-field considerations above.

Let us first consider the master equation for the number of active neurons $A^\mathrm{rest}$ within the non-driven population.
Assuming asynchronous transitions, $A^\mathrm{rest}$ can change at most by one: It can increase through activation of a previously inactive neuron with rate $\Omega^\mathrm{rest}_+$; it can decrease through inactivation of a previously active neuron with rate $\Omega^\mathrm{rest}_-$; or it can not change through reactivation of an already active neuron or deactivation of an already inactive neuron with rate $\Omega^\mathrm{rest}_0$. 
The transition rates for each neuron depend, in general, on how many of its pre-synaptic neurons from the driven and non-driven population are active. 
To specify the transition rates, we assume that the rate to activate any single neuron per unit time is given by the mean probability of recurrent activation, Eq~\ref{eq_mean-field_prob}, which can be rewritten as $\overline{p^\mathrm{rec}}=\lambda\left[A^\mathrm{in} + A^\mathrm{rest}\right]/N$ and thus depends on both $A^\mathrm{in}$ and $A^\mathrm{rest}$.
Since single neuron activation are independent, transition rates sum up and we arrive at
\begin{align}\label{eq:rates_rest}
    \Omega^\mathrm{rest}_+(A^\mathrm{rest}) &=  (N^\mathrm{rest}-A^\mathrm{rest})\overline{p^\mathrm{rec}}\nonumber\\
    \Omega^\mathrm{rest}_-(A^\mathrm{rest}) &=  A^\mathrm{rest}\left(1-\overline{p^\mathrm{rec}}\right)\\
    \Omega^\mathrm{rest}_0(A^\mathrm{rest}) &=  A^\mathrm{rest}\overline{p^\mathrm{rec}} + (N^\mathrm{rest}-A^\mathrm{rest})\left(1-\overline{p^\mathrm{rec}}\right) \nonumber
\end{align}
The master equation that describes the time evolution of the probability $P(A^\mathrm{rest},t)$ to find the non-driven population in state $A^\mathrm{rest}$ at time t, is then
\begin{align*}
\dot{P}(A^\mathrm{rest},t) =
\phantom{+}&\Omega^\mathrm{rest}_+(A^\mathrm{rest}-1)P(A^\mathrm{rest}-1,t)\nonumber\\
+&\Omega^\mathrm{rest}_-(A^\mathrm{rest}+1)P(A^\mathrm{rest}+1,t)\\
-&\left(\Omega^\mathrm{rest}_+(A^\mathrm{rest})+\Omega^\mathrm{rest}_-(A^\mathrm{rest})\right)P(A^\mathrm{rest},t),\nonumber
\end{align*}
where contributions from $\Omega^\mathrm{rest}_0(A^\mathrm{rest})$ cancel out. 

From this master equation, we can obtain a Fokker-Planck equation through a Kramers-Moyal expansion where we keep only terms up to second order~\cite{risken_fokker-planck_2012}:
\begin{align}
\frac{\partial P(A^\mathrm{rest},t)}{\partial t}
=-&\frac{\partial \left(r^\mathrm{in}\left(A^\mathrm{rest}\right) P \left(A^\mathrm{rest},t\right)\right)}{\partial A^\mathrm{rest}}\label{eq_fokker_planck_rest}\\
+&\frac{1}{2} \frac{\partial^2\left(g^\mathrm{in}\left(A^\mathrm{rest}\right)P\left(A^\mathrm{rest},t\right)\right)}{\partial^2 A^\mathrm{rest}}\nonumber,
\end{align}
with drift coefficient $r^\mathrm{rest}(A^\mathrm{rest})=\Omega^\mathrm{rest}_{+}(A^\mathrm{rest})-\Omega^\mathrm{rest}_{-}(A^\mathrm{rest})$ and diffusion coefficient $g^\mathrm{in}(A^\mathrm{rest})=\Omega^\mathrm{rest}_{+}(A^\mathrm{rest})+\Omega^\mathrm{rest}_{-}(A^\mathrm{rest})$.

Analogously, we can obtain a Fokker-Planck equation for the distribution of active neurons $P(A^\mathrm{in},t)$ within the input population. 
Here, neurons can be independently activated by external input with probability $p^\mathrm{ext}=1-e^{-h\Delta t}$, resulting in the transition rates
\begin{align}
    \Omega^\mathrm{in}_+(A^\mathrm{in}) &=  (N^\mathrm{in}-A^\mathrm{in})\left(\overline{p^\mathrm{rec}}e^{-h\Delta t}+1-e^{-h\Delta t}\right)\nonumber\\
    \Omega^\mathrm{in}_-(A^\mathrm{in}) &=  A^\mathrm{in}\left(1-\overline{p^\mathrm{rec}}\right)e^{-h\Delta t}
    \label{eq:rates_in}
\end{align}
with a Fokker-Plank equation as in \eqref{eq_fokker_planck_rest} but with drift coefficient $r^\mathrm{in}(A^\mathrm{in})=\Omega^\mathrm{in}_{+}(A^\mathrm{in})-\Omega^\mathrm{in}_{-}(A^\mathrm{in})$ and diffusion coefficient $g^\mathrm{in}(A^\mathrm{in})=\Omega^\mathrm{in}_{+}(A^\mathrm{in})+\Omega^\mathrm{in}_{-}(A^\mathrm{in})$.

Recall, however, that the transition rates $\overline{p^\mathrm{rec}}=\overline{p^\mathrm{rec}}(A^\mathrm{in}, A^\mathrm{rest})$ depend on both $A^\mathrm{in}$ and $A^\mathrm{rest}$, such that the resulting Fokker Planck equations are coupled.
Since we cannot solved these coupled equations, we seek an approximate solution after decoupling them. 
To decouple the equations, we assume for each Fokker Planck equation that the other population can be described by its mean-field equation. 
To be specific, consider the Fokker Planck equation for $A^\mathrm{rest}$.
Then, we can obtain from Eq.~\eqref{eq_mean-field_equation_in} the self-consistent equation $A^\mathrm{in}=a^\mathrm{in}\mu N=1-(1-\lambda(A^\mathrm{in}+A^\mathrm{rest})/N)e^{-h\Delta t}$, which we can solve for $A^\mathrm{in}=f(A^\mathrm{rest})$. 
We repeat the same procedure from Eq.~\ref{eq_mean-field_equation_rest} to obtain $A^\mathrm{rest}=g(A^\mathrm{in})$.
When we enter these back into the equation for $\overline{p^\mathrm{rec}}$, we obtain
\begin{align*}
    \overline{p^\mathrm{rec}}(A^\mathrm{rest}) \approx& 
   \frac{ A^\mathrm{rest}}{N}\frac{\lambda}{1-\lambda\mu e^{-h\Delta t}} + \lambda\mu\left(1-e^{-h\Delta t}\right) 
    \\
    \overline{p^\mathrm{rec}}(A^\mathrm{in}) \approx&
    \frac{A^\mathrm{in}}{N}\frac{\lambda}{1-(1-\mu)\lambda}
\end{align*} 
These can then be used to fully specify the transition rates, Eq.~\eqref{eq:rates_rest} and Eq.~\eqref{eq:rates_in}, respectively, and thereby the corresponding drift and diffusion terms in the Fokker Planck equations for $P(A^\mathrm{rest},t)$ and $P(A^\mathrm{in},t)$.

The decoupled Fokker Planck equations can be solved assuming stationary dynamics, i.e., $\partial P/\partial t = 0$. 
Using change of variables, it is straightforward to show that the stationary state is described by the probability distribution~\cite{risken_fokker-planck_2012}
\begin{equation}
    P(x)\approx \frac{1}{g(x)}\exp\left\{2\int\frac{r(x)}{g(x)}dx\right\},
\end{equation}
where $x=\{A^\mathrm{in}, A^\mathrm{rest}\}$ is the activity in the respective population.
In a simplified notation, we can write $r(x)= a x + c_0$ and $f(x)= b x^2 + dx + c$, where the coefficients \{$a,b,c,d\}$ only depend on the constants $\{\lambda, \mu, h, \Delta t, N\}$.
Then, we obtain the solutions
\begin{widetext}
\begin{equation}
    	  P(x) \sim \\
    	  (bx^2+dx+c)^{\frac{a}{b}-1} \exp\left( \frac{(ad-2bc)}{b\sqrt{d^2-4bc}} \log(\frac{\sqrt{d^2-4bc}+(2bx+d)}{\sqrt{c^2-4bc}-(2bx+d)}, \right),
    	\label{eq:fokker_planck_solution}
\end{equation}
\end{widetext}
which belong to the exponential family. 
Since we are interested in the distribution of the output response, which we here assume is the sum of all active neurons, i.e., $A=A^\mathrm{in}+A^\mathrm{rest}$, we further need a convolution over both decoupled solutions and obtain 
\begin{equation}
    P(A) = (P^\mathrm{in}\ast P^\mathrm{rest})(A).
    \label{}
\end{equation}
Due to the functional form of the $P(x)$, this convolution is not analytically tractable.
We thus perform the computations numerically to obtain a semi-analytic approximation.
Note that if $N^\mathrm{out} = N$, we can obtain a semi-analytical solution for the distribution of outputs for a single time step $P(o^T)$, with $T=1$, by substituting $A=No^T$.

\begin{figure*}
    \centering
    \includegraphics[width=1\textwidth]{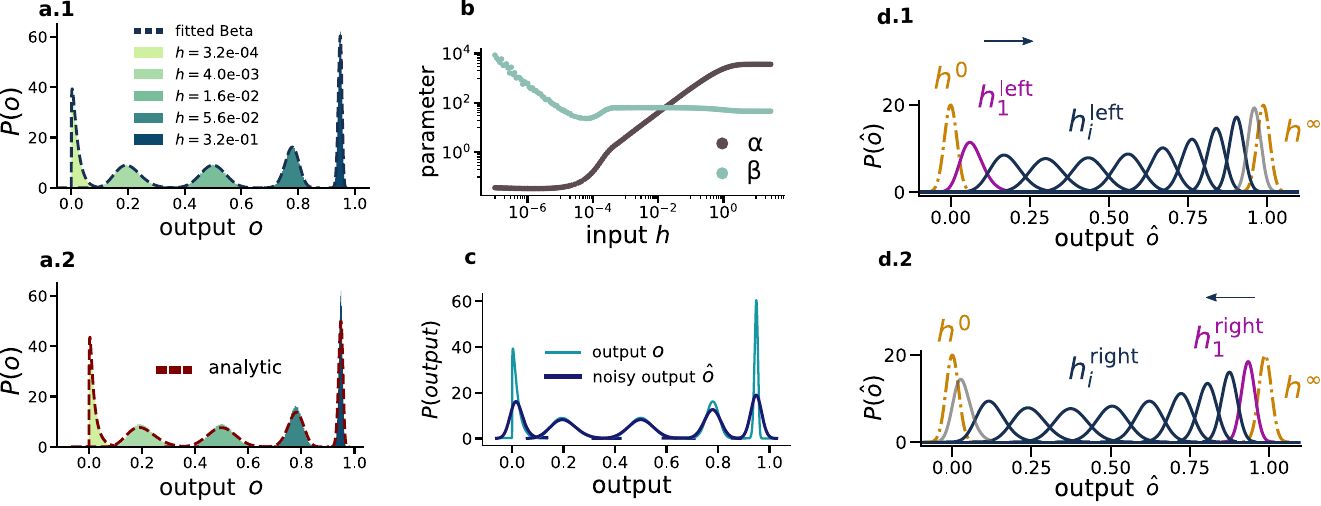}
    \caption{Workflow to determine processing capacity.
    We start from a continuum approximation of the clean output response. While this is a direct result of the analytical calculations in \textbf{a.2}, for the numerical results we 
    \textbf{a.1} approximate the distributions by Beta distributions and for analytical predictions
    \textbf{b} interpolate the parameters of the Beta distributions in between sampled input values.
    \textbf{c} We obtain estimates for the distribution of noisy outputs for a given $h$ from a convolution with a Gaussian distribution.   
    \textbf{d.1} and \textbf{d.2} We determine the smallest (largest) discriminable input as the first input value for which the distribution $P(\hat{o}|h)$ overlaps with the minimal (maximal) active state by less than the discrimination error $\varepsilon$.
    From this we obtain an estimate of the dynamic range.
    We further determine the number of discriminable inputs as those for which all distributions overlap by less than $\varepsilon$.
    Here we used $\lambda=0.9968$, $\varepsilon=0.1$ and $\sigma_{noise}=0.02$ for better visualisations.}
    \label{fig:extData_workflow}
\end{figure*}
\endgroup

\clearpage
\renewcommand{\thefigure}{S\arabic{figure}}
\setcounter{figure}{0}
\renewcommand{\thetable}{S\arabic{table}}
\setcounter{table}{0}

\onecolumngrid
\section*{Supplementary Information}

\subsection{Mean-Field solution for the network response}
We can write the mean-field approximation for the responses of the network. We denote the response of the input population as the fraction of the active neurons $a^{in}=n/N^{in}$ and the response of the rest of the network as $a^{rest}=m/N^{rest}$ where $n$ and $m$ are the numbers of active neurons in the two populations. The response of the input population is given by
\begin{equation}
    a^\mathrm{in}=1-\left(1-\mathbb{E}(p^\mathrm{rec})\right)\left(1-p^\mathrm{ext}\right)\,
    \label{app1_eq1}
\end{equation}
where the average activation probability from recurrence in the network is given by
\begin{equation}
    \mathbb{E}(p^\mathrm{rec})=\lambda\left(\mu a^\mathrm{in}+ \left(1-\mu\right)a^\mathrm{rest}\right)\,
    \label{recurrentInput}
\end{equation}

and $\mu=N^\mathrm{in}/N$ is the fraction of neurons that are receiving the input. The response of the rest of the population can be computed as
\begin{equation}
    a^\mathrm{rest}= \mathbb{E}\left(p^\mathrm{rec}\right)
    \label{app1_eq2}.\
\end{equation}
We can solve the two equations \ref{app1_eq1} and \ref{app1_eq2} to obtain the responses as a function of the external input rate h given that $p^\mathrm{ext}=1-e^{-h\Delta t}$. Namely,

\begin{equation}
    a^\mathrm{in}=\frac{1-e^{-h\Delta t}}{1-\frac{\lambda \mu e^{-h\Delta t}}{1-\lambda \left(1-\mu\right)}}\,
\end{equation}

\begin{equation}
    a^\mathrm{rest}=\frac{\lambda \mu (1-e^{-h\Delta t})}{1-\lambda\left(1-\mu\right)-\lambda\mu^{-h\Delta t}}
\end{equation}

\subsection{Derivation of response distribution }
We can consider the activation and deactivation of the neurons as a one-step death-and-birth process where we consider an asynchronous update of the network state (i.e. updating the states of the neurons one by one). We show that the response distribution derived from this master equation is identical to the dynamics of our network with synchronous update.\\

We start by considering the input population with $n(t)$ active nodes at time $t$. For simplicity of the notation, we drop the functionality in the rest of the text.
The rate of transitioning from $n$ active units to $n-1$ is $\Omega^{in}(n\to n-1):=\Omega^{in}_{-}$ that can be written in terms of average input to each neuron from other neurons  $\mathbb{E}(p_{rec})$ as

\begin{equation}
    \Omega^\mathrm{in}_{-}=n\left(1-\mathbb{E}\left(p^\mathrm{rec}\right)\right)\left(1-p^\mathrm{ext}\right),
    \label{omega-in}
\end{equation}
The rate of the opposite transition $\Omega^\mathrm{in}(n \rightarrow n+1):=\Omega^\mathrm{in}_{+}$ can be written as
\begin{equation}
    \Omega^\mathrm{in}_{+}=(N^\mathrm{in}-n)\left(\mathbb{E}\left(p^\mathrm{rec}\right)+\left(1-\mathbb{E}\left(p^\mathrm{rec}\right)\right)p^\mathrm{ext}\right).
    \label{omega+in}
\end{equation}
Now we can write the master equation as
\begin{equation}
    \dot{P}\left(n,t\right)=\Omega^\mathrm{in}_{+}P\left(n-1,t\right)+\Omega^\mathrm{in}_{-}P\left(n+1,t\right)-\left(\Omega^\mathrm{in}_{+}+\Omega^\mathrm{in}_{-}\right)P\left(n,t\right).
    \label{master1}
\end{equation}

One convenient way to solve this Master equation is to use the Kramers-Moyal~\cite{} expansion and keep the terms up to the second order to obtain the approximate Fokker-Planck equation. The corresponding Fokker-Planck equation is 
\begin{equation}
    	\frac{\partial P(n,t)}{\partial t}=-\frac{\partial 
    	\left(r^\mathrm{in}\left(n\right) P \left(n,t\right)\right)}{\partial n}+\frac{1}{2} \frac{\partial^2 \left(g^\mathrm{in}\left(n\right)P\left(n,t\right)\right)}{\partial^2 n},
    	\label{FPE1}
\end{equation}
where the drift and diffusion coefficients are $r^\mathrm{in}(n)=\Omega^\mathrm{in}_{+}(n)-\Omega^\mathrm{in}_{-}(n)$ and $g^\mathrm{in}(n)=\Omega^\mathrm{in}_{+}(n)+\Omega^\mathrm{in}_{-}(n) $ respectively.

Similarly we can write the transition rates for the non-driven population with $m$ active nodes as follows
\begin{equation}
    \Omega^{-}_{m}=m\left(1-\mathbb{E}\left(p^\mathrm{rec}\right)\right),
    \label{omega-m}
\end{equation}

\begin{equation}
    \Omega^{+}_{m}=\left(N^\mathrm{rest}-m\right)\mathbb{E}\left(p^\mathrm{rec}\right).
    \label{omega+m}
\end{equation}

The corresponding FPE then would be
\begin{equation}
    	\frac{\partial P(m,t)}{\partial t}=-\frac{\partial (r^\mathrm{rest}\left(m\right)P\left(m,t\right))}{\partial m}+\frac{1}{2} \frac{\partial^2 \left(g^\mathrm{rest}\left(m\right)P\left(m,t\right)\right)}{\partial^2 m},
    	\label{FPE2}
\end{equation}
where 
$r^\mathrm{rest}(n)=\Omega^\mathrm{rest}_{+}(n)-\Omega^\mathrm{rest}_{-}(n)$ and $g^\mathrm{rest}(n)=\Omega^\mathrm{rest}_{+}(n)+\Omega^\mathrm{rest}_{-}(n) $ respectively.

Note that all transition rates and therefore drift and diffusion coefficients are functions of both $m$ and $n$, because $\mathbb{E}(p^\mathrm{rec})=\lambda\left(\mu n/N^\mathrm{in}+ \left(1-\mu\right)m/N^\mathrm{rest}\right)$ which couples equation eq\ref{FPE1} and eq~\ref{FPE2}. There is no general way to analytically solve two coupled partial differential equations with high order polynomial coefficients. In order to obtain an approximation of the solution we decouple the two equations by substituting the mean field equations for $n$ and $m$ in the drift and diffusion coefficients. Namely, substituting $m=\lambda (1-\mu) n /\left(1-\left(1-\mu\right)\lambda\right)$ (derived from equation \ref{app1_eq2}) into $g^\mathrm{in}$ and $r^\mathrm{in}$. Similarly we substitute $n=\left(N\mu p^\mathrm{ext}-\lambda \mu \left(1-p^\mathrm{ext}\right)m\right)/\left(1-\lambda \mu \left(1-p^\mathrm{ext}\right)\right)$ (derived from equation \ref{app1_eq1}) into coefficients $g^\mathrm{rest}$ and $r^\mathrm{rest}$.
This allows us to write the drift and diffusion coefficients as polynomials. We have
\begin{equation}
    r^{i}=A^{i} x+C^{i},
    \label{r_in}
\end{equation}
\begin{equation}
    g^{i}=B^{i} x^2+D^{i} x+ C^{i},
    \label{g_in}
\end{equation}
where $i\in \{in,rest\}$ and $x$ is the continous representative of $n$ ($i=in$ ) and $m$ ($i=rest$). Now we can solve the decoupled FPEs  obtain the probability distribution of stationary state ($\frac{\partial P}{\partial t} =0$). With a change of variable it is then straightforward to show that the stationary state has a probability distribution  
\begin{equation}
    	P(x)\sim \frac{1}{g^{i}(x)}\exp\left(2 \int \frac{r^{i}(x)}{g^{i}(x)} \,dx\right).
    	\label{prob}
\end{equation}
The probability distribution has the final form of 
\begin{equation}
    	  P(x) \sim \\
    	  (B^{i}x^2+D^{i}x+C^{i})^{\frac{A^{i}}{B^{i}}-1} \exp\left( \frac{(A^{i}D^{i}-2B^{i}C^{i})}{B^{i}\sqrt{{D^{i}}^2-4B^{i}C^{i}}} \log(\frac{\sqrt{{D^{i}}^2-4B^{i}C^{i}}+(2B^{i}x+D^{i})}{\sqrt{{C^{i}}^2-4B^{i}C^{i}}-(2B^{i}x+D^{i})}, \right)
\end{equation}
where the coefficients ,$A^{i}$, $B^{i}$, $AC^{i}$ and $D^{i}$ depend on only parameters of the network architecture, namely $N$, $\lambda$, $\mu$ and $T$. Note that the probability distribution belongs to the exponential family. Since we're only interested in the effect of parameters $\lambda$ and $T$on the behavior of the distributions we only keep these two and absorb the rest of the parameters in the form of the functions.Therefore we get
\begin{equation}
        n \sim f^\mathrm{in}(\lambda,T) 
\end{equation}
and
\begin{equation}
        m \sim f^\mathrm{rest}(\lambda,T),
\end{equation}
where $f^\mathrm{in}$ and $f^\mathrm{rest}$ are the probability density functions. 
In order to derive the probability distribution of the total number of active neurons $k=n+m$ we convolve the two density functions. Therefore we obtain the probability density function of k as 
\begin{equation}
  u(k)= (f^\mathrm{in}*f^\mathrm{rest})(k).
\end{equation}
Since the functional form of $f$ is complicated, and therefore the convolution not analytically tractable, we perform the computations numerically.

\subsection{Readout Gaussian noise}
We additionally consider that independent from the observation time, the readout circuitry might be intrinsically noisy and we include this noise as an additive source in the model. Therefore we have
\begin{equation}
    Y = K+ Z ,\
    \label{eq15}
\end{equation}
where $Y_T$ is the noisy output and $\eta$ is the noise term that for simplicity is considered Gaussian\\
\begin{equation}
    Z\sim \mathcal{N}(0,\,\sigma_{noise}^{2}).\
    \label{eq16}
\end{equation}
The probability distribution of the output $Y$ can be obtained simply via the convolution of the  distribution of the readout $f^\mathrm{out}(k)$ and the noise $g_z$. Namely, 
\begin{equation}
  q(y)= \Phi(\left(u_k*g_{z}\right))(y),\
\end{equation}
where function $\Phi$ makes sure that the support of $q_{Y_T}$ stays in the meaningful range of $[0,1]$. This is done by adding the probability of those events outside this range to the probability at 0 and 1.

\subsection{Computing discrimination error}
 To measure the inference error we consider two inputs $h_{1}$ and $h_{2}$ and their corresponding output response probability distributions $P({\hat{O}|h_1})$ and $P({\hat{O}|h_2})$. The overlap between the two probability distributions quantifies the Minimal Discrimination Error \cite{berens2009neurometric} of an ideal observer in classifying the input ~\cite{berens2009neurometric}. Namely, 
 \begin{equation}
   \mathcal{E}(h_1,h_2)=\frac{1}{2}\int min\left(P({\hat{o}|h_1}),P({\hat{o}|h_2})\right)\left(\hat{o}\right) \,d\hat{o} .
    \label{eq13}
\end{equation}
We calculate the overlaps numerically.

\subsection{Finding discriminable inputs}
To obtain the set of discriminable inputs, we start by identifying the smallest, reliably discriminable input from the zero-active state ($h \to 0$). 
In the limit of $h \to 0$, the stationary activity (for $\lambda\leq1$) is a Dirac delta function at zero since that is an absorbing state in the absence of input. Therefore the noisy output for all $T$ will be a Gaussian distribution centered at zero.
 Therefore the smallest discriminable input $h_1^{left}$ must satisfy
\begin{equation}
    \mathcal{E}(h=0,h_1^{left})=\varepsilon ,\
    \label{dscrInp1}
\end{equation}
where $\varepsilon$ is our acceptable error threshold. The superscript ``left'' is used to highlight that we started off by finding the smallest input (left-hand side in the response curve) and will make our way up to larger values.
We calculate $h_1^{left}$ using an optimization algorithm to find the root of eq\ref{dscrInp1}. We can find the next discriminable inputs in a similar manner by solving the equation
\begin{equation}
    \mathcal{E}(h_j^\mathrm{left},h_{j+1}^\mathrm{left})=\varepsilon ,\
    \label{dscrInp2}
\end{equation}
We denote this set of inputs as $\{h_i^{left}\}$.
A similar procedure can be taken by identifying the largest input that can be reliably discriminable from the output in the limit $h \to \infty$ (which can be found the lowest value above which network activity is saturated.) For the current network architecture, the maximally active state does not correspond to $\hat{o}=1$ since only a fraction of the network receives the input. $h_1^{right}$ must satisfy
\begin{equation}
    \mathcal{E}(h_1^\mathrm{right},h \to \infty)=\varepsilon ,\
    \label{dscrInp3}
\end{equation}
and all the next discriminable inputs are frond by solving 
\begin{equation}
    \mathcal{E}(h_{j-1}^\mathrm{right},h_j^\mathrm{right})=\varepsilon .\
    \label{dscrInp4}
\end{equation}
This yield the second set $\{h_i^{right}\}$.
We can then compute the dynamic range as 
\begin{equation}
   \Delta= 10\log_{10}\left(\frac{h_{1}^{right}}{h_{1}^{left}}\right).\
    \label{DR}
\end{equation}
We define inference resolution as the arithmetic mean of the number of discriminable inputs in both sets $\{h_i^{left}\}$ and $\{h_i^{right}\}$. Namely, 
\begin{equation}
   res=\frac{N_d}{\Delta}
    \label{eq20}
\end{equation}
where
\begin{equation}
   N_d=\frac{1}{2}\left(\#\{h_i^{left}\}+\#\{h_i^{right}\}\right),
    \label{N_d}
\end{equation}
and $\#\{\cdot\}$ denotes a cardinality (number of elements) of a set.

\begin{figure*}
    \centering
    \includegraphics[width=1\textwidth]{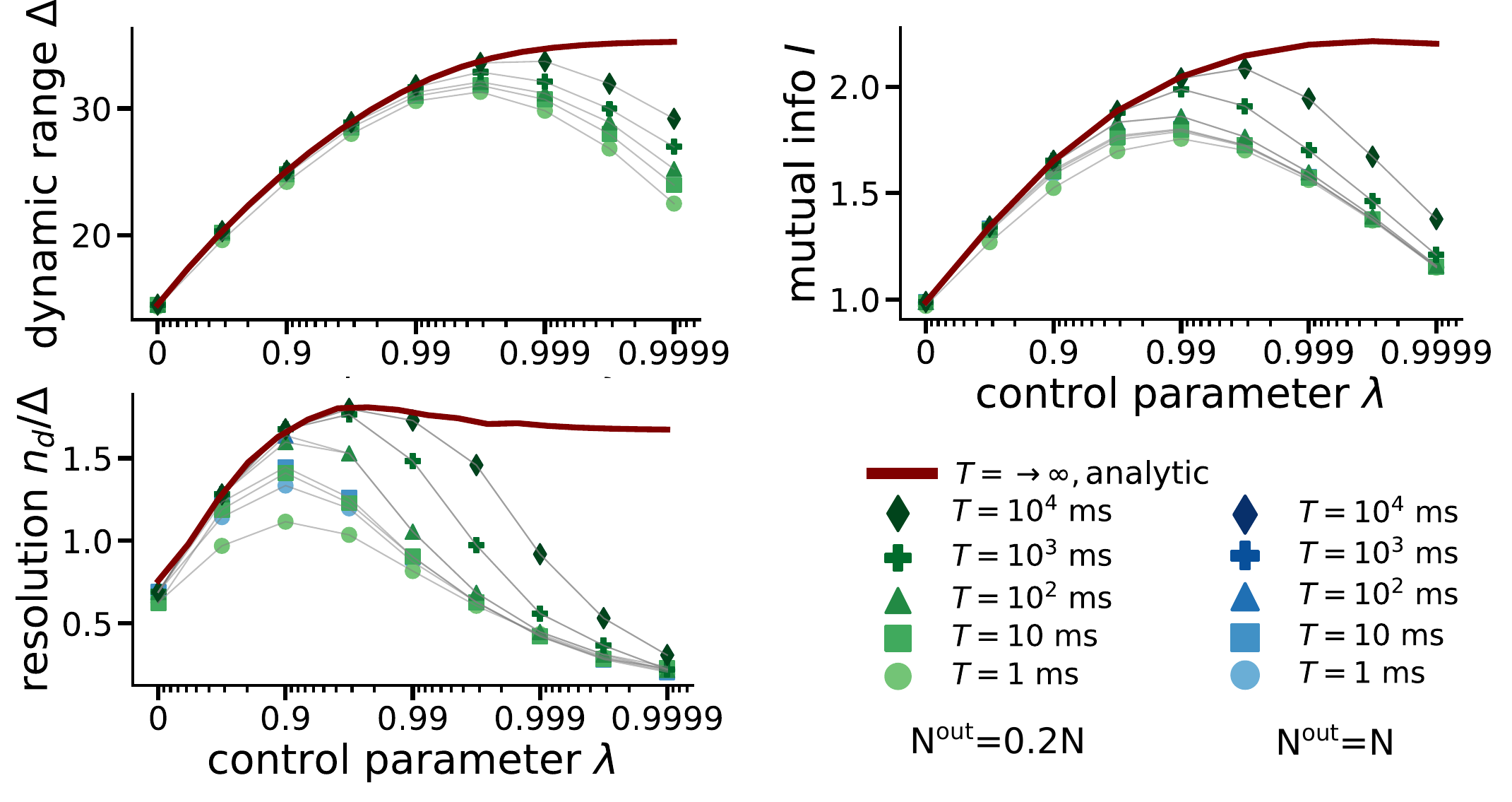}
    \caption{\fs{The main results remain qualitatively unchanged when the output neurons are a subset (subsample) of the whole population.} Comparison between the two case of $N^\mathrm{out}=N$ and $N^\mathrm{out}=0.2N$ for \textbf{a)} Dynamic range, \textbf{b)} inference resolution and \textbf{c)} Mutual information between input and mean output activity given different observation windows. Parameter values are $N^\mathrm{in}=2000$, $N=10000$, $\varepsilon=0.2$ and $\sigma_{noise}=0.01$.
    }
    \label{supp2}
\end{figure*}

\begin{figure*}
    \centering
    \includegraphics[width=1\textwidth]{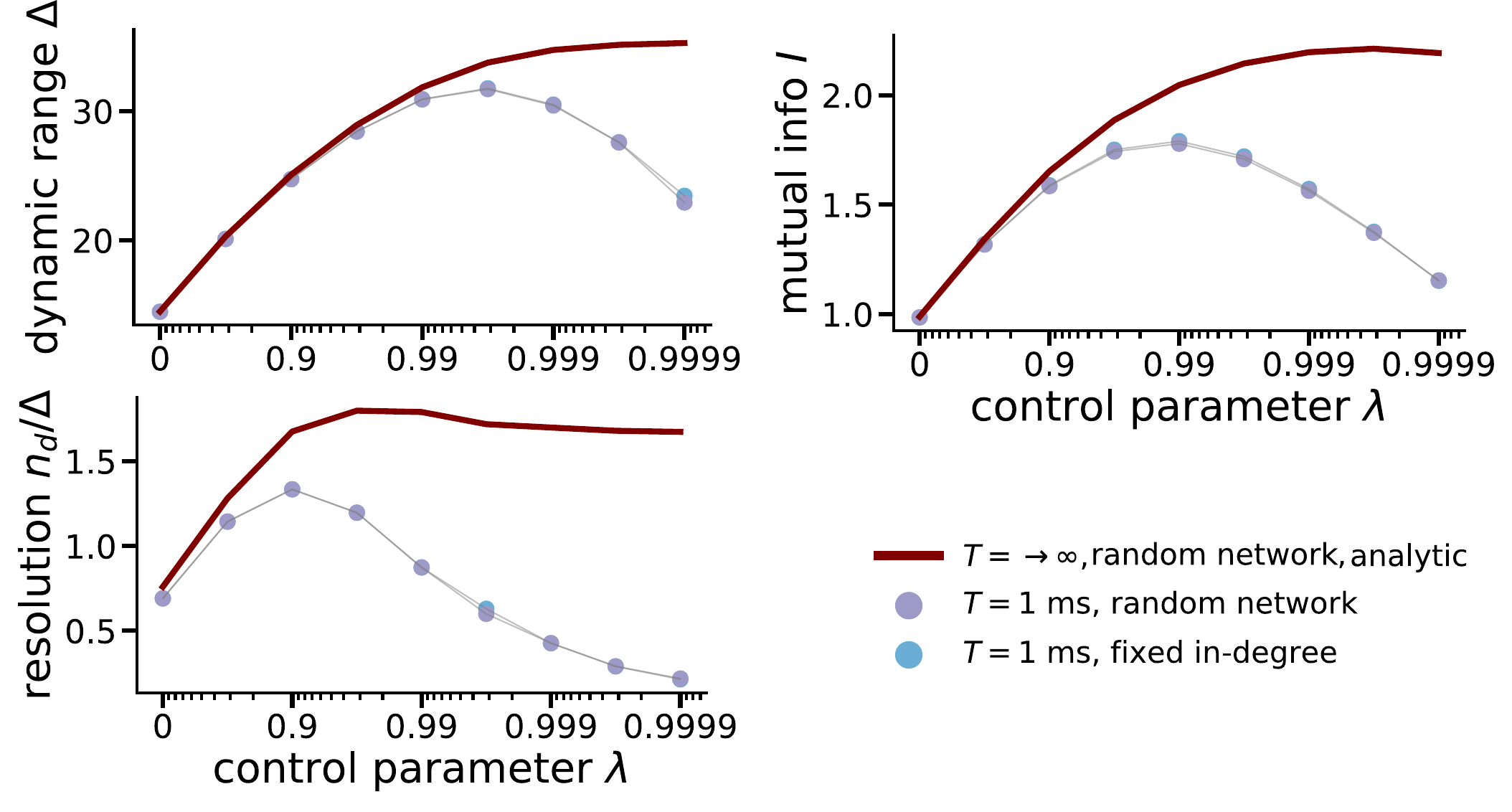}
    \caption{\fs{ The main results generalize to a random network.} Comparison between a network with random connections and the one used in the main paper with fixed in-degree\textbf{a)} Dynamic range, \textbf{b)} inference resolution and \textbf{c)} Mutual information between input and mean output activity given T=$1$ms.  Parameter values are $N^\mathrm{in}=2000$, $N^\mathrm{out}=10000$, $\varepsilon=0.2$ and $\sigma_{noise}=0.01$.}
    \label{supp3}
\end{figure*}

\subsection{Analysis workflow}
\fs{1. Simulation of the network dynamics} \\

To obtain the output distributions we simulate the evolution of the network in time.
We generate the synaptic weight Matrix $W$ using networkx library. For simplicity we chose a regular network (equal in-degree connectivity of K for all neurons). In order to speed up the computations we chose $K=100$ and equal weights of $w=\lambda/K$ for all neurons. Here $\lambda$ is the control parameter and is in the range $[0,1)$. We use scipy.sparse library to construct the matrix.\\

The state of neurons are updated in two stages in every time step. First, They can be activated by the recurrent input from withing the network with a probability $p^\mathrm{rec}[s_i(t+\Delta t)=1]=f(\sum_j w_{ij} s_j(t))$ (which is implemented by a matrix product). If the "input" neuron is not activated at this step, it still has a chance to become active via an external input with probability $p^\mathrm{ext}[s^\mathrm{in}_i(t+\Delta t)=1]=1-e^{-h\Delta t}$. All the neurons that did not become active after these two steps, will be silenced irrespective of their previous state. This implies a refractory period of one timestep. We consider each time-step to last around a millisecond since the refractory period of neurons is generally about one milliseconds .\\

The network response at each time-point is computed as $a^\mathrm{out}(t)=\frac{1}{N^\mathrm{out}}\sum_i s^\mathrm{out}_i(t)$ which is the average firing rate of the output population. The output of the network is computed by averaging the network response over a given processing window $T$ as $o_T=\frac{1}{T}\sum_t a^\mathrm{out}(t)$. \\

We simulate the network dynamics for as long as $10^8$ timesteps. Moreover, we only keep the stationary part of the time series (the first $10^4$ time steps are discarded). We divide the timeseries into non-overlapping chunks of size $T\in  {1,10,10^2,10^3,10^4}$ and compute the output by averaging the response in those windows.This yields minimum $10^4$ output values (and more for smaller window sizes). \\

We run these simulations for $9$ values of $\lambda$ logarithmically spaced in the range $[0,0.9999]$. These compuatations are repeated for $170$ values of input intensities logarithmically spaced the range $[10^{-7},10^{1.5}]$. Additionally, in order to compute the error bars in our measurements we repeat the simulation for $10$ networks with different weight matrix $W$. 
\\

\fs{2. Obtaining output distributions}\\
We either construct the output distribution using the analytical solution for the two limits $T\in 1ms$ and $T \in \infty$ (see methods) or by Fitting Beta distributions and interpolation in the parameter space.\\

\fs{2.1 Using analytical solution}\\
To obtain the analytical form of the distribution at $T=1ms$ We normalize the analytical solution derived from solving the FPE (see methods). Since it is tedious to perform analytical operations on the function in it's original form, we do all the computations (such as normalization, convolution, etc.) numerically. We use Sympy library for the symbolic evaluations of the probability distribution.\\

The analytical form of the output distribution at $T\to \infty$ is equivalent to a Dirac delta function with its mean value calculated from Eq.~\eqref{eq:mean}. We do not actively use this form in our computations as it always appear as a Gaussian distribution after being corrupted by the readout noise (see 3.).\\

\fs{2.2 Fitting Beta distribution and interpolating in the parameter space}\\
In order to obtain output distribution for any values of input, we do a linear interpolation of the distribution parameters by approximating the data with a (standard) Beta distribution. The Beta distribution is fitted to the output data using rv\_continuous.fit class from scipy.stats library which uses a Maximum Likelihood Estimation. Figure ~\ref{fig:extData_workflow}a shows how well the Beta distribution describes the output distribution given a variety of input values. \\

Standard Beta distribution (defined over $[0,1]$) has two  shape parameter $\alpha$ and $\beta$. Figure ~\ref{fig:extData_workflow}c shows the parameter landscape from which we can linearly interpolate $\alpha(h)$ and $\beta(h)$ for any value of $h$ within the defined range.

\fs{3.Adding Gaussian noise}\\
Another source of noise that we consider is the readout Gaussian noise that is additively added to the output, $\hat{o}=o+\eta$ where $\eta$ comes from a Gaussian distribution with mean zero and variance $\sigma_\mathrm{noise}$. Therefore, the probability of the noisy output $\hat{o}$ is computed via the convolution of the distribution of the two distributions i.e,
\begin{equation}
   P(\hat{o})=(P_o*\phi_{0,\sigma^2_\mathrm{noise}})(\hat{o}),
    \label{noise1}
\end{equation}
where $\phi_{0,\sigma^2_{noise}}$ is the Gaussian distribution and $P_o$ is the raw output distribution which either comes from 2.1 or 2.2.\\

The convolution is done with two arrays of $10^4$ sampled data points from the two distribution using the convolve function from numpy library. Figure~\ref{fig:extData_workflow}d shows a comparison between the original and the noisy output distributions for a given value of input.  \\

Note that after this point $P(o)$ is used in the form of a sampled array of size $10^4$. \\

\fs{4.Finding the smallest and largest discriminable inputs and dynamic range}\\
To determine the smallest discriminable input we find the the input $h_1^\mathrm{left}$ that satisfies
\begin{equation}
    \int \min\left(P\left(\hat{o}|h_1^\mathrm{left}\right ),P\left(\hat{o}|h\to 0\right)\right)\left(\hat{o}\right) \,d\hat{o}-2\varepsilon=0 ,
     \label{workFlow2}
\end{equation}
where $P(\hat{o}|h\to0)$ is just a Gaussian distribution with mean zero and variance $\sigma_\mathrm{noise}$ i.e $\phi_{0,\sigma_\mathrm{noise}}$.\\

To determine the largest discriminable input we find the input $h_1^{\mathrm{right}}$ that satisfies
\begin{equation}
    \int \min\left(P\left(\hat{o}|h_1^\mathrm{right}\right ),P\left(\hat{o}|h\to\infty\right)\right)\left(\hat{o}\right) \,d\hat{o}-2\varepsilon=0 ,
     \label{workFlow3}
\end{equation}
where $P\left(\hat{o}|h\to\infty\right)$ is obtained either from 2.1 or 2.2. Note that since the network activity saturates above a certain value of input, we can find the saturated distribution withing the available range of $h$.\\

The integral is computed numerically using quad function from scipy.integrate library.\\

Next, We find $h_1^\mathrm{left}$ and $h_1^\mathrm{left}$ by finding the root  of eq~\ref{workFlow2} in the $h$ dimension, using bisect function from scipy.optimize library.\\
Figure~\ref{fig:extData_workflow}e shows the zero and maximally active states in gray, and the smallest and largest discriminable inputs in magenta.\\
Finally dnamic range is computed as $10log(h_1^\mathrm{left}/h_1^\mathrm{right})$.
\\

\fs{6.Finding all discriminable inputs and inference resolution}\\
We can find all the discriminable inputs iteratively starting from the smallest to the largest i.e to solve the following equation
\begin{equation}
    \int \min\left(P\left(\hat{o}|h_{i+1}^\mathrm{left}\right ),P\left(\hat{o}|h_i^\mathrm{left}\right)\right)\left(\hat{o}\right) \,d\hat{o}-2\varepsilon=0 .
\end{equation}
The equation is solved similar to section 4 and $i$ goes from $1$ (using $h_1^\mathrm{left}$ from previous section) to ${n_d}^{left}$, after which the equation has no solution anymore.\\

We do the same procedure starting from the largest discriminable input and iteratively to the smallest one i.e 
\begin{equation}
    \int \min\left(P\left(\hat{o}|h_{i+1}^\mathrm{right}\right ),P\left(\hat{o}|h_i^\mathrm{right}\right)\right)\left(\hat{o}\right) \,d\hat{o}-2\varepsilon=0 .
\end{equation}
This time $i$ goes from $1$ (using $h_1^\mathrm{right}$ from previous section) to ${n_d}^{right}$.\\

Lastly, we compute the inference resolution as  $n_d=\left(n_d^\mathrm{left}+n_d^\mathrm{right}\right)/2\Delta$.
\\

\end{document}